\documentclass[%
 reprint,
 amsmath,amssymb,
 aps,superscriptaddress,
floatfix,
nofootinbib
]{revtex4-2}

\usepackage{graphicx}
\usepackage{dcolumn}
\usepackage{amsmath}
\usepackage{bm}
\usepackage{csquotes}
\usepackage{enumitem}
\usepackage{booktabs}
\usepackage[dvipsnames]{xcolor}
\definecolor{convection_color}{HTML}{AD339C}
\definecolor{collision_color}{HTML}{B9B4AF}
\definecolor{transport_color}{HTML}{30B9BB}
\definecolor{relaxation_color}{HTML}{E77335}
\definecolor{external_force_color}{HTML}{318909}
\usepackage[colorlinks=false, hidelinks]{hyperref}
\usepackage{bookmark}

\begin{document}

\title{A collisional model of odd fluids: from Boltzmann equation to chiral hydrodynamics}
\author{Ege Eren}
\affiliation{James Franck Institute, University of Chicago, Chicago, IL 60637, USA}

\author{Michel Fruchart}
\affiliation{Gulliver, CNRS, ESPCI Paris, Université PSL, 75005 Paris, France}

\author{Vincenzo Vitelli}
\affiliation{James Franck Institute, University of Chicago, Chicago, IL 60637, USA}
\affiliation{Leinweber Institute for Theoretical Physics, University of Chicago, Chicago, IL 60637, USA}

\def \restcoefnormal{r_\parallel} 
\def \restcoeftangent{r_\bot}
\def \modtangentfric{e_\bot}
\def \apsevector{\hat{a}}
\def \fullcollop{\mathcal{C}}
\def \linearcollop{C}
\def \torquecollop{L_\omega}
\def \boltzmannop{L}

\begin{abstract}
When the time-reversal and parity symmetries in a fluid are broken, transverse transport coefficients can arise in response to perturbations, an example being odd viscosity. We refer to these systems as odd fluids. While much progress has been made in the continuum theory of odd-viscous fluids, and non-collisional models for odd viscous fluids have been proposed, a classical microscopic description in which the transverse responses originate from collisions is lacking. In this paper, we show that a dilute granular gas of rough and inelastic particles driven by a constant torque is a minimal microscopic model of an odd fluid. By applying the methods of Boltzmann kinetic theory, we obtain a hydrodynamic description of the microscopic model. Then, using the method of adiabatic elimination, we numerically compute all the response coefficients of the model, explicitly showing that the model has many odd response terms. Our theory predicts that certain odd response coefficients can change sign even when the direction of the external torque is fixed. While we choose a particular case, the procedure we present can be applied to any collisional model. We also present a semi-quantitative method to determine the hydrodynamic variables of the theory by observing the eigenvalue spectrum of the linear collision operator.


\end{abstract}

\maketitle


\section{Introduction}
\label{sec:level1}

Chiral fluids are made of individual elements that spin, breaking both time-reversal and mirror symmetries.
These systems exhibit transverse responses to perturbations, known as odd or Hall responses, that must be accounted for in a hydrodynamic description that goes beyond the standard Navier-Stokes equations describing usual fluids~\cite{fruchart2023odd,Avron1998}. 
Examples include systems such as polyatomic gases \cite{beenakker1970magnetic,korving1966transverse,korving1967influence,hulsman1970experimental} and plasma \cite{braginskii1958transport,berdyugin2019measuring} under a magnetic field, as well as rotating gases~\cite{nakagawa1956kinetic}.
More recently, active chiral fluids have been proposed and experimentally realized in soft matter platforms \cite{banerjee2017odd,fruchart2023odd,Liebchen2022,LopezCastano2022,Scholz2018}.
Experiments~\cite{Soni2019} and numerical simulations~\cite{han2021fluctuating,hargus2020time,Hargus2021,Lou2022} have indeed shown that odd responses arise in fluids of spinning objects.
In order to understand and predict the properties of chiral fluids, it is desirable to connect their large-scale hydrodynamic description to microscopic models through a coarse-graining procedure.

At the microscopic level, odd responses can originate from two classes of mechanisms: (i) the behavior of a single (composite) particle, and (ii) chiral collisions between particles. 
In the context of active matter, most work on coarse-graining up to now has focused on the first class of mechanisms~\cite{markovich2021odd,markovich2022nonreciprocity,matus2024molecular,Poggioli2023} while neglecting the effect of collisions.
Exceptions include Refs.~\cite{Ghimenti2023,Ghimenti2024,Ghimenti2024b}, in which mode coupling theory is used to compute the response in systems with transverse forces added to speed up the convergence to thermal equilibrium, as well as Refs.~\cite{fruchart2022odd,Jiao2024}, where kinetic theory is used to coarse-grain systems with simplified collision models.
Based on the results of these works, as well as numerical simulations~\cite{han2021fluctuating}, and comparison with magnetized polyatomic gases~\cite{beenakker1970magnetic}, we expect that collisional effects can be responsible for a substantial part of the odd responses.

The goal of this work is then to derive a hydrodynamic theory describing the large-scale behavior of a collisional active chiral fluid starting from a realistic microscopic model. 
In order to do so, we start with a minimal but realistic model of a two-dimensional granular gas where inelastic, rough disks are actively driven by a constant single-body torque (Sec. \ref{microscopic_description}). 
The combination of external torques and interparticle friction drives the granular gas out of equilibrium, leading to a pseudo-thermal nonequilibrium stationary state (Sec. \ref{statistical_description}).
Using kinetic theory, we derive the hydrodynamic equations describing the corresponding compressible chiral fluid and calculate the generalized response tensors including odd response coefficients such as odd viscosity and odd thermal conductivity (Sec. \ref{sec:hydordynamics}). By dimensional analysis, we show that all the response tensors are proportional to the square root of the constant external field.
We numerically compute all the relaxation and transport coefficients for different values of friction parameters (Sec. \ref{numerical_computation}).
We observe that the response coefficients can change sign as these friction parameters are varied, even if the direction of the external field is kept the same.
Our results suggest that for certain values of the microscopic parameters, a predictive hydrodynamic description of compressible odd fluids may require to take into account additional slow variables in the hydrodynamic equations in addition to conserved quantities.

\section{Microscopic description}
\label{microscopic_description}

\subsection{Microscopic model}

We consider a two-dimensional dilute gas of hard disks with radius $a$ and mass $m$, that evolve under a constant driving torque $\vec{d} = d \vec{e}_z$, where $\vec{e}_z$ is the out-of-plane unit vector. 
Unless otherwise specified, we assume that the gas made of $N$ disks is contained in a box of size $L$ with periodic boundary conditions, so the average density is $n=N/L^2$.

Each disk is described by the position of its center of mass $\vec{r}_i$, the velocity of the center of mass $\vec{c}_i$, and the angular velocity of the disk $\vec{\omega}_i = \omega_i \vec{e}_z$.
The evolution of the system is described by Newton's equations
\begin{equation}
    m \frac{d^2 \vec{r}_i}{dt^2} = m \frac{d \vec{c}_i}{dt} = \sum_j \vec{f}_{j \to i}
    \label{resultant}
\end{equation}
and
\begin{eqnarray}
    I \frac{d \vec{\omega}_i}{dt} = \vec{d} + \sum_j \vec{\tau}_{j \to i}
    \label{moment}
\end{eqnarray}
in which $\vec{f}_{j \to i}$ and $\vec{\tau}_{j \to i}$ are inter-particle forces and torques (relative to the center of mass of disk $i$), respectively, and where $I = q ma^2$ with $q=1/2$ is the moment of inertia of a disk.

The effect of interparticles forces and torques in Eqs.~\eqref{resultant} and \eqref{moment} is modeled by collisions between the disks, that we describe in detail in the next section.
The numerical results reported in this paper are obtained using this collision model, but our general strategy can be applied to other situations, provided that interactions can be summarized by well-defined collisions that are separated from each other. 
In particular, we have assumed the gas is dilute enough that collisions between three or more particles can be neglected, so only binary collisions are taken into account.



\subsection{Description of binary collisions}

In order to describe collisions, we consider a model of rough hard disks commonly used to describe granular media~\cite{Goldsmith2001,Jenkins1985,Lun1991,Lun1987,Walton1993,Brilliantov1996,Zippelius2006,Brilliantov2010,Poschel2008}.
The key point of this model is that transverse frictional forces arise at the point of contact between colliding rough disks, coupling translational and rotational degrees of freedom and dissipating energy.

\begin{figure}
\includegraphics{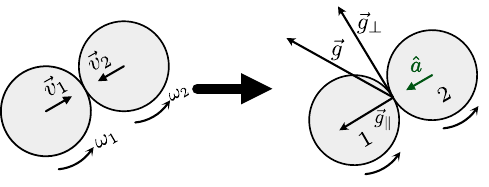}
\caption{ \textbf{Microscopic Pair-Collision of Disks with Normal and Tangential Friction:} Depiction of a collision in the reference frame of particle 2. $\apsevector$ denotes the apse vector and $\Vec{g} = \Vec{c}_{1} - \vec{c}_2 + \apsevector\times(\Vec{\omega}_1 + \omega_2)$ is the relative velocity of the point of contact.}
\label{fig:pair_collision}
\end{figure}

The collision model can be summarized as follows. Suppose two particles with velocities $(\vec{c}_{1,2}, \vec{\omega}_{1,2} )$, collide with a hard-core interaction (Fig.~\ref{fig:pair_collision}). If we denote the unit vector from center of the second particle to the point of contact as $\apsevector$ (also known as the apse vector \footnote{ In general, the apse vector is the unit vector from the center of mass of particle 2 to the trajectory of particle 1, where the two particles are the closest to each other. For hard sphere collisions, this turns out to be the point of contact.}) and define $\Vec{g} = \vec{c}_1 - \vec{c}_2 + a \apsevector \times (\vec{\omega}_1 + \vec{\omega}_2)$ as the relative velocities of particles at the point of contact, the post-collision relative velocity satisfy
\begin{equation}
    \Vec{g}^{\,\prime}_\parallel = -\restcoefnormal \Vec{g}_\parallel
    \quad
    \text{and}
    \quad
    \Vec{g}^{\,\prime}_\perp = \restcoeftangent \Vec{g}_\perp.
\end{equation}
where
\begin{eqnarray}
    \Vec{g}_\parallel = (\apsevector \cdot \Vec{g}) \apsevector
    \quad
    \text{and}
    \quad
    \Vec{g}_\perp = \Vec{g} - \Vec{g}_\parallel
\end{eqnarray}
denote the component of $\Vec{g}$ that are respectively parallel and perpendicular to $\apsevector$, and in which the parameters $ \restcoefnormal $ and $\restcoeftangent$ are constant restitution coefficients (see Refs.~\cite{Zippelius2006, Brilliantov2000, Brilliantov2010} for discussions on the effect of velocity-dependent restitution coefficients).

Imposing the conservation of total linear momentum and the conservation of angular momentum at the point of contact, post-collision (denoted by primed symbols) velocities of particle 1 and particle 2 are found to be
\begin{subequations}
\begin{align}
    \label{eq:post-vel1}
    \vec{c}_1^{\,\prime} &= \vec{c}_1 - \frac{1+\restcoefnormal}{2}\vec{v}_n - \modtangentfric(\Vec{v}_t+\Vec{v}_r)
    \\
    \label{eq:post-vel2}
    \vec{c}_2^{\,\prime} &= \vec{c}_2 + \frac{1+\restcoefnormal}{2}\vec{v}_n + \modtangentfric(\Vec{v}_t+\Vec{v}_r)
    \\
    \label{eq:post-rot-vel}
    \Vec{\omega}_{1,2}^{\,\prime} &= \Vec{\omega}_{1,2} + \frac{\modtangentfric}{a q}[\apsevector\times(\Vec{v}_t + \Vec{v}_r)],
\end{align}
\end{subequations}
where
\begin{subequations}
\begin{align}
    \Vec{v}_n &= [(\vec{c}_1 - \vec{c}_2)\cdot\apsevector] \, \apsevector
    \\
    \Vec{v}_t &= (\vec{c}_1 - \vec{c}_2) - \Vec{v_n}
    \\
    \Vec{v}_r &= a \apsevector\times(\Vec{\omega_1} + \Vec{\omega_2})
\end{align}
\end{subequations}
and $\modtangentfric \equiv q(1-\restcoeftangent)/(2q+2)$. 
Note that this collision rule allows linear momentum to be exchanged with angular momentum for each particle. 
Nevertheless, each collision conserves the total linear momentum, as Eqs.~(\ref{eq:post-vel1}-\ref{eq:post-vel2}) imply that
\begin{equation}
    m\vec{c}_1^{\,\prime} + m \vec{c}_2^{\,\prime} = m \vec{c}_1 + m \vec{c}_2.
\end{equation}
In contrast, energy is not conserved.

\section{Statistical description}
\label{statistical_description}

\subsection{Boltzmann equation}

To describe the statistical properties of our chiral granular gas, we turn to the Boltzmann equation
\begin{equation}
    \label{eq:Boltzmann_eq}
        \partial_t f + \vec{c}\cdot\nabla f + d \partial_\omega f = \mathcal{C}(f,f),
\end{equation}
which describes the evolution of the single particle probability distribution function $f(\Vec{r},\vec{c},\vec{\omega},t)$, where  $\vec{c}$ is the single particle center of mass velocity, ${\omega}$ is the single particle angular velocity, $d$ is the external torque and $\mathcal{C}(f,f)$ is a collision operator accounting for the effect of collisions \cite{dorfman2021contemporary,chapman1990mathematical}.
The collision operator corresponding to our model is
\begin{equation}
\label{eq:collision_operator}
\begin{split}
  \mathcal{C}(f,g)[\vec{r},t,\vec{c}_1,\omega_1] = 2a \int d\apsevector d^2 c_2 d\omega_2 
  \Theta(-\vec{c}_{12}\cdot\apsevector) 
  \\
  \times |\vec{c}_{12}\cdot\apsevector|
  \; \left[ \frac{1}{\beta r^2}f_1''g_2'' - f_1 g_2 \right],
\end{split}
\end{equation}
which is derived in detail in App. \ref{appendix:coll_operator}.
Here $''$ denotes the pre-collision quantities that result in two particles with velocities $(\vec{c}_1,\omega_1)$ and $(\vec{c}_2,\omega_2)$ respectively and $f_n = f(t, \vec{r}_n,\vec{c}_n,\omega_n)$. We also introduced the shorthand notation $\vec{c}_{12} \equiv \vec{c}_1 - \vec{c}_2$.
The Boltzmann equation relies on the molecular chaos hypothesis (\emph{Stosszahlansatz}), according to which particles about to collide are completely uncorrelated before the collision. 
This requires the chiral gas to be sufficiently dilute. 

While we have considered a 2D gas of homogeneous hard disks ($q = 1/2$), all the procedure described until now is equally applicable to 3D spheres by promoting all the vectors to 3D and choosing the parameter $q$ appropriately (for instance for homogeneous hard spheres $q = 2/5$).

\subsection{Non-equilibrium steady-state}
\label{sec:quasi-equilibrium}

A granular gas in which no energy is injected from outside quickly stops moving because all its kinetic energy is dissipated by friction \cite{Brilliantov2010,Poschel2008}. 
In contrast, a granular system submitted to an external drive (such as a vibrating wall) can reach a non-equilibrium steady-state where energy injection and dissipation compensate. 
We refer to Refs.~\cite{Zippelius2006} and \cite[ch.~9]{dorfman2021contemporary} for a review on the kinetic theory description of both of these cases.

The presence of external torques combined with interparticle friction in the model described in Sec.~\ref{microscopic_description} can in principle lead to a non-equilibrium steady-state. 
Indeed, the energy injected in the rotational degrees of freedom by the external torque $\vec{d}$ can be transferred into the translational degrees of freedom through collisions. 
At the same time, both rotational and translational energies are lost by friction during collisions.
This suggests that the gain and loss may be balanced in a non-trivial steady-state. 
Such a steady-state $f_{\text{ss}}$, if it exists, is a solution of the stationary homogeneous Botzmann equation
\begin{equation}
    d \partial_\omega f_{\text{ss}} = \fullcollop(f_{\text{ss}},f_{\text{ss}}).
    \label{shbe}
\end{equation}
We expect, based on numerical simulations (see below and Refs.~\cite{han2021fluctuating}) that a non-trivial steady-state solution of Eq.~\eqref{shbe} exists, but finding its exact analytical expression is not straightforward. 
We instead consider a Gaussian ansatz of the form
\begin{equation}
    \label{eq:equilibrium_distribution}
    f(\Vec{c},\vec{\omega},t) = \mathcal{N}(t) \, \exp{\left[-\frac{m \lVert\vec{c}\, \rVert^2}{2 T_t(t)} - \frac{I \lVert \vec{\omega} - \vec{\omega}_0(t) \rVert^2}{2 T_r(t)} \right]},
\end{equation}
with
\begin{eqnarray}
    \mathcal{N}(t) = n \frac{m}{2\pi T_t(t)} \sqrt{\frac{I}{2\pi T_r(t)}},
\end{eqnarray}
where $n$ is the particle density, $\omega_0$ is the average angular velocity and $T_t$ and $T_r$ are the translational and rotational temperatures, respectively. 
The validity of this Gaussian ansatz is discussed in App.~\ref{app:MD-sim}, in which we present the results of a small scale molecular dynamics simulations that validate our approach. 
In short, the simulations show that the distributions of linear velocities are close to Gaussian at high normal restitution coefficients $\restcoefnormal $, but deviate from Gaussian at low $\restcoefnormal$, while the distribution of angular velocities always exhibits an asymmetry (see Refs.~\cite{puglisi2002fluctuation, cafiero2000two,goldhirsch2005nearly} for similar situations). 
Nevertheless, numerical simulations of similar systems suggest that such a Gaussian approximation captures the mean features of the model at the level of the observables considered here~\cite{han2021fluctuating}. 
This is similar to non-chiral granular gases, in which non-Gaussian corrections can also be obtained perturbatively starting from the Gaussian approximation \cite{Zippelius2006}. 
%

The evolution in time of the parameters within the ansatz \eqref{eq:equilibrium_distribution} is then described by the equations
\begin{subequations}
\label{homogeneous_ds}
\begin{align}
    \!G^{-1}\dot{\omega}_0 &= G^{-1} d - \frac{C_1}{2} {\omega}_0 T_t^{1/2}  
    \\
    \!G^{-1}\dot{T_t} &= -A_1 T_t^{3/2} + A_2 T_t^{1/2} T_r + A_2 T_t^{1/2}[2I{\omega}_0^2] 
    \\
    \!G^{-1}\dot{T_r} &= 2 A_2 T_t^{3/2} - 2 B_1 T_r T_t^{1/2} + 2 B_2 T_t^{1/2}[2I{\omega}_0^2]
\end{align}
\end{subequations}
obtained from the Boltzmann equation \eqref{eq:Boltzmann_eq} by multiplication with $\omega/n$, $m c^2/(2 n)$, ${I(\vec{\omega}-\omega_0)^2}/n$, respectively, and integration over phase space.
The overdot represents time derivatives.
We have defined
\begin{subequations}
\begin{equation}
    G = \frac{8 a n \sqrt{\pi}}{\sqrt{m}}
    \quad
    A_1 = \left[  \frac{1- \restcoefnormal ^2}{4} + \frac{\modtangentfric(1-\modtangentfric)}{2}   \right]
\end{equation}
as well as
\begin{equation}
    A_2  =  \frac{\modtangentfric^2}{2q}
    \quad
    B_1 = \frac{\modtangentfric}{2q}\left( 1 - \frac{\modtangentfric}{q} \right)
\end{equation}
and
\begin{equation}
    B_2 = \frac{\modtangentfric^2}{2q^2}
    \quad
    C_1 = \frac{2\modtangentfric}{q}
\end{equation}
\end{subequations}

Fixed points of the dynamical system \eqref{homogeneous_ds} can be found by solving the three algebraic equations corresponding to $\dot\omega_0 = \dot T_t = \dot T_r = 0$. 
This gives
\begin{subequations}
\begin{align}
    \label{eq:omega_equilibrium}
    \omega_0 &= \frac{2 d}{G C_1 T_t^{1/2}}\\
    \label{eq:rot_T_equilibrium}
    T_r &= \frac{A_1}{A_2} T_t - \frac{8 q d^2 m a^2}{G^2 C_1^2 T_t}\\
    \label{eq:tr_T_equilibrium}
    T_t &= \sqrt{m} a |d| \left[  \frac{\frac{8q}{C_1^2 G^2} (B_2 + B_3) }{   \frac{B_2 A_1}{A_2} - A_2 } \right]^{1/2}
\end{align}
\end{subequations}
which we will define to be the average equilibrium angular velocity $\omega_0$, the rotational temperature $T_r$ and translational temperature $T_t$, respectively. 
A linear stability analysis of \eqref{homogeneous_ds} shows that this fixed point is stable (App.  \ref{app: Stability Analysis}), confirming that it is indeed an approximately equilibrium state.
Note that the only parameter that sets the time scale in this system is the external torque and there is no external reference energy scale.

While we will focus on the Gaussian approximation of the steady-state distribution in the rest of the paper, we emphasize that our method does not rely on this hypothesis, which is needed only to evaluate the integrals. 
The Gaussian approximation could be replaced by any analytical approximation or a numerical solution of the homogeneous stationary Boltzmann equation if greater accuracy is desired in lieu of analytical tractability. 


\section{Hydrodynamics}
\label{sec:hydordynamics}

Our goal is to obtain a hydrodynamic description of the chiral active gas, similar to the Navier-Stokes equations in standard fluids, starting from the statistical description of Sec.~\ref{statistical_description} and using kinetic theory to perform coarse-graining~\cite{dorfman2021contemporary,chapman1990mathematical}.
In order to do so, we proceed in three steps.

First, we need to identify the relevant hydrodynamic variables to be included in the coarse-grained description. 
In a nutshell, the goal is to identify a set of variables approximately describing the state of the system and that are predictive of their own future through a closed set of coupled differential equations, or more generally through an effective field theory.
These variables should be physical observables, that are represented by functions $\chi(\vec{c},\vec{\omega})$ on phase space.
The macroscopic average of an observable $\chi$ is then defined as
\begin{equation}
    \label{eq:mic-to-mac}
     \rho_\chi(\vec{r},t)
     \equiv 
     \langle \chi (\vec{r},t)\rangle
     \equiv
     \frac{1}{n(\vec{r},t)}\int d\vec{c} d\omega f(\vec{c},\omega,\vec{r},t) \chi(\vec{c},\omega)
 \end{equation}
in which $n(\vec{r},t)$ is the density defined as
\begin{equation}
    n(\vec{r},t) \equiv \int d \vec{c} d\omega f(\vec{c},\omega,\vec{r},t)
\end{equation}
In order to select the relevant observables for the hydrodynamic description, we rely on a separation of scales between slowly-decaying variables (to be kept in the hydrodynamic description) and quickly-decaying ones (to be integrated out) \footnote{Note that in the Gaussian ansatz \eqref{eq:equilibrium_distribution}, we have already guessed that the hydrodynamic or almost-hydrodynamic variables are density, linear and angular momenta, as well as translational and rotational kinetic energies. 
This is a consequence of our inability to solve Eq.~\eqref{shbe} directly. 
In principle, assuming that the response stays linear, one can first solve Eq.~\eqref{shbe} and then use the linearized Boltzmann equation (linearized about the solution of Eq.~\eqref{shbe}) to identify the slow (hydrodynamic) variables.}.

Second, we can obtain the equation of change for $\rho_\chi$ (also known as equation of transfer) from the Boltzmann equation \eqref{eq:Boltzmann_eq} by multiplying it by $\chi(\vec{c},\vec{\omega})$ and integrating over  phase space, see \cite[\S~3.13]{chapman1990mathematical}. 
The equations obtained in this way are exact (under the assumption that the Boltzmann equation holds), but they are not closed. 
Instead, several additional variables (such as the pressure tensor, the heat flux vector, and so on) can be identified. 
Their value in a uniform steady-state can be determined from the results of Sec.~\ref{sec:quasi-equilibrium} and encoded in equations of state.  

Third, we use the linear response approximation in order to determine the values of the additional variables when the hydrodynamic variables are not uniform. 
We obtain the corresponding linear response coefficients, known as relaxation and transport coefficients, by using adiabatic elimination on the linearized Boltzmann equation, i.e. by tracing out all the degrees of freedom that are not included in the hydrodynamic fields.


\subsection{Linearized Boltzmann Equation}
\label{lbe}

We first analyze the fate of small perturbations $\phi$ about the stationary distribution $f^{(0)}$ \footnote{We will denote the steady-state Gaussian ansatz as $f^{(0)}$ and any numerical result given below is computed using such ansatz. However, the techniques presented for the rest of the paper do not really depend on the form of the steady-state distribution function, and any step can be reproduced if a suitable orthonormal basis is chosen under the inner product defined with another steady-state distribution $f_{ss}.$} that denotes the solution of the homogeneous Boltzmann equation \eqref{eq:Boltzmann_eq}, that is, we consider $f = f^{(0)}(1 + \phi)$, where $|\phi(\Vec{r},\vec{c},\vec{\omega},t)| \ll 1$. 
Keeping only the linear terms in $\phi$, we find the linearized Boltzmann equation
\begin{equation}
\label{eq:lin_Boltzmann_eq}
    \partial_t \phi + \vec{c}\cdot \nabla \phi + d\partial_\omega \phi + d \phi \partial_\omega (\log f^{(0)}) = C\phi,
\end{equation}
in which the linearized collision operator $C$ is defined by
\begin{align}
    C\phi(\vec{r},t,\vec{c}_1,\omega_1) &= \frac{1}{f_1^{(0)}}\left[ \mathcal{C}( f^{(0)}, f^{(0)}\phi) + \mathcal{C}( \phi f^{(0)}, f^{(0)} )\right] \nonumber \\
    &= \frac{2a}{f_1^{(0)}}\int d\apsevector d\vec{c}_2 d\omega_2 |\vec{c}_{12}\cdot\apsevector| \Theta(\vec{c}_{12}\cdot\apsevector) \nonumber \\
    &\times \left[ \frac{f_1''^{(0)} f_2''^{(0)} }{\beta r^2}(\phi_1'' + \phi_2'') - f_1^{(0)} f_2^{(0)}(\phi_1 + \phi_2) \right]
\end{align}
where lower indices $(1,2)$ indicate functions evaluated with the velocities $(\vec{c}_{1,2}, \omega_{1,2})$ of particles $(1,2)$.

A spatial Fourier transform of Eq.~\eqref{eq:lin_Boltzmann_eq} yields
\begin{equation}
    \label{eq:lin_Boltzmann_eq_Fourier}
    \partial_t \phi + i \vec{q}\cdot\vec{c} \, \phi = L \phi,
\end{equation}
where we have defined the Boltzmann operator $L$ as
\begin{equation}
    L \equiv C - d\partial_\omega - d\partial_\omega(\log f^{0}) .
\end{equation}
The wavevector $\vec{q}$ can be seen as an inhomogeneity parameter; the long-wavelength regime of almost spatially uniform perturbations corresponds to the limit $\vec{q} \to 0$.

To bridge the Boltzmann equation with the hydrodynamic description, we define the inner product
\begin{equation}
    \label{eq:inner-product}
    (\chi,\zeta) \equiv \frac{1}{n}\int d\vec{c}_1 d\omega_1 f^{(0)}_1 \bar{\chi} \zeta
\end{equation}
for any functions $\zeta$ and $\chi$ on phase space. 
Following Eq.~(\ref{eq:mic-to-mac}), we can then associate an equilibrium macroscopic average $\rho_\chi^0$ with the inner product $\rho_\chi^0 = (\chi,1)$ and the perturbation $\delta \rho_\chi$ on the equilibrium distribution with $\delta \rho_\chi = (\chi,\phi)$, where $\phi(\vec{r},\vec{c},\omega,t)$ is the small perturbation function on the Boltzmann distribution defined above. 

With all these definitions, one can then obtain the linearized macroscopic equation of change for any variable $\chi(\vec{c},\omega)$ by taking the inner product of equation \eqref{eq:lin_Boltzmann_eq_Fourier} with $\chi$, which yields
\begin{equation}
    \label{evolution_perturbation}
    \partial_t \delta \rho_\chi + iq\cdot(\chi, \vec{c}) = \left( \chi, L \phi \right) = (L^\dagger \chi,\phi).
\end{equation}
As the form of the equation suggests, we define a collision invariant as a function $\chi$ that satisfies $(L^\dagger\chi,\phi) = 0$ for any function $\phi$.
Equivalently, this means that $\chi$ is in the nullspace of the adjoint Boltzmann operator $L^\dagger$.

In the long-wavelength limit ($q \to 0$), the perturbation $\delta \rho_\chi$ along an eigenfunction $\chi$ of $L^\dagger$ with eigenvalue $\lambda_\chi$ behaves as
\begin{equation}
    \rho_\chi(t) \simeq e^{\lambda_\chi t} \rho_\chi(0),
\end{equation}
which quickly relaxes to zero.
Hence, only the nullspace of $L^\dagger$ and, possibly, eigenfunctions with a small enough eigenvalue contribute to  the long wavelength behavior. 
These correspond to the hydrodynamic variables of the theory.

We emphasize that the Boltzmann operator is not Hermitian ($L^\dagger \neq L$), which can be traced to the combination of friction and drive. 
As can be seen from Eq.~\eqref{evolution_perturbation}, it is $L^\dagger$ that is relevant to define the hydrodynamic variables, rather than $L$. 
In Sec. \ref{hydrodynamic_variables} below, we present a numerical example where the eigenfunctions and eigenvalues of $L^\dagger$ are computed.

\begin{figure}
\includegraphics[scale=0.8]{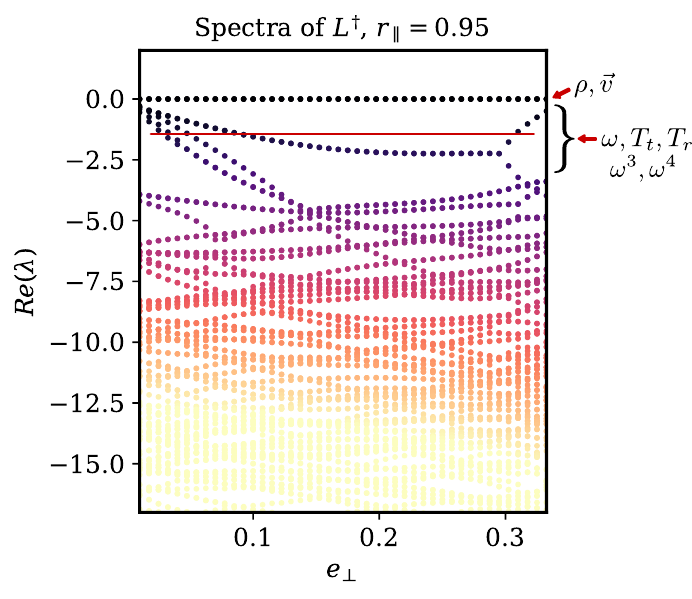}
\caption{\label{fig:epsart} \textbf{Eigenvalue Spectra of the Hermitian Conjugate of the Boltzmann Operator and the Separation of Slow Variables:} Eigenvalue spectra of the Hermitian conjugate of the linearized Boltzmann operator for a fixed normal restitution coefficient $\restcoefnormal = 0.95$ across various tangential restitution $\modtangentfric$, in dimensionless units. As can be seen, when the friction is sufficiently small, the eigenvalues that do not correspond to the conserved quantities can be close to 0 and should be considered as hydrodynamic variables. As the tangential friction is increased, the gap between collision invariants and temperatures and rotational velocity increases. Beyond very small $\modtangentfric$ values, the hydrodynamic variables remain as a linear combination of angular momentum $(\omega - \omega_0)$, and translational and rotational temperatures $T_t$ and $T_r$. When $\modtangentfric$ is increased even further, the only eigenvalue that remains near the vicinity of the zero modes still turns out to be a linear combination of these variables and none of the conventional variables separate. The red line is completely arbitrary and is inserted to give a general idea of our approach to the hydrodynamic variables.}
\label{fig:collision_operator_eigenval}
\end{figure}

\subsection{Spectrum of the linearized collision operator and hydrodynamic variables}
\label{hydrodynamic_variables}

Figure~\ref{fig:collision_operator_eigenval} shows a numerical approximation of the eigenvalue spectrum of $L^\dagger$ with a fixed normal restitution coefficient to $\restcoefnormal = 0.95$ and different values of tangential restitution $\modtangentfric$ (see Sec.~\ref{numerical_computation} and App. ~\ref{app:Computational_Details} for details on the numerics). 
The eigenvalues with largest real part (growth/decay rates) are shown.
We first observe that all (real parts of the) eigenvalues are nonpositive, i.e. there is no unstable eigenvalue. 
In addition, zero is an eigenvalue, and it is always three-fold degenerate when $\modtangentfric \neq 0$. 
Indeed, conserved quantities belong to the nullspace of $L^\dagger$, and here we can check numerically that they generate the nullspace, which is spanned by $1$ and $c_i$ ($i=x,y$).
All the other eigenvalues correspond to decay. 
We also observe that for certain ranges of $\modtangentfric$, there is a gap in the eigenvalue spectrum that separates groups of eigenvalues.
This spectral separation corresponds to a separation of time scales between the corresponding eigenvectors and serves as a basis for determining the hydrodynamic variables.
Namely, we consider as hydrodynamic variables the averages constructed from the observables encoded in the eigenvectors with decay rates above the gap, while non-hydrodynamic variables are all the other (quickly decaying) degrees of freedom.

The number of variables to take into account depends on the degree of approximation made. For very low tangential restitution, the eigenvalues that correspond to angular momentum, translational and rotational temperatures, and higher powers of angular momentum remain very close to zero-modes. In the more intermediate regime, the higher powers of the angular velocity $O(\omega - \omega_0)^3$ separate and move toward the fast variable region and considering mass, linear and angular momenta, and translational and rotational temperatures is sufficient. When the variable $\modtangentfric$ is increased even further, we are left with three or five slow variables, depending on where the cutoff value is chosen, the first three being the conserved quantities. The fourth and fifth slow variables turn out to be linear combinations of the angular momentum, and translational and rotational temperatures. 

For the rest of the paper, and in the explicit calculations, we take the hydrodynamic variables to be the density, linear and angular momenta, and translational and rotational temperatures.

\subsection{Equations of change}
\label{equations_of_change}

Once the slow variables are determined, the corresponding equations of change can be obtained by multiplying the Boltzmann equation \eqref{eq:Boltzmann_eq} by the corresponding variables and integrating them over the phase space with the recipe given in equation \eqref{eq:mic-to-mac}.
We consider mass density $\rho$, linear momenta $\rho \vec{u}$, angular momentum $In\bar{\omega}$, translational kinetic energy $e_t$ and rotational kinetic energy $e_r$, respectively defined through the microscopic observables as
\begin{subequations}
\begin{align}
    \rho \equiv m n \\
    u_i \equiv \langle c_i \rangle\\
    \omega_0 \equiv \langle \omega \rangle\\
    n e_t \equiv \langle m\vec{c}^2/2 - m \vec{u}^2/2 \rangle, \\
    n e_r \equiv \langle I\omega^2/2 - I \omega_0^2/2 \rangle.
\end{align}
\end{subequations}
We also define the associated temperatures as $e_t = T_t$ and $e_r = \frac{1}{2}T_r$, where we set $k_B = 1$. Using these definitions and applying the steps described above, we find the corresponding equations of change \cite[\S~3.13]{chapman1990mathematical}
\begin{subequations}
\label{equations_of_change}
\begin{align}
    D_t \rho &= - \rho \nabla \cdot \vec{u} \\
    \rho D_t u_i &= - \partial_j P_{ij} + nb_i \\
    In D_t \Bar{\omega} &= - \partial_i M_i + Idn + \mathcal{C}[I\omega]\\
    nD_t e_t &= - \nabla \cdot \Vec{Q} - P_{ij} \partial_i u_j + \mathcal{C}[mv^2/2]\\
    \label{eq:full_hydro2}
    nD_t e_r &= - \nabla \cdot \Vec{Q}_\omega - M_i \partial_i \Bar{\omega} + \mathcal{C}[I\omega^2/2 - I\omega\Bar{\omega}]
\end{align}
\end{subequations}
where we have defined
\begin{equation}
    C[\psi] \equiv \int d\vec{c}_1 d\omega_1 \mathcal{C}(f,f) \psi(\vec{c}_1,\omega_1)
\end{equation}
which vanishes if $\psi$ is a collision invariant, see App. \ref{appendix:coll_operator} for more details. 
Here, $\vec{b}$ is the external force that in general can be included, but that we set to zero. 
We have furthermore introduced quantities representing the fluxes of the hydrodynamic quantities, namely defined the pressure tensors (fluxes of linear and angular momentum)
\begin{equation}
    \label{eq:pressure_tensor}
    P_{i j} = \rho\langle U_i U_j \rangle
    \quad
    \text{and}
    \quad
    M_i = I n \langle U_i W \rangle
\end{equation}
as well as the heat flux vectors (fluxes of translational and rotational kinetic energy)
\begin{equation}
    \label{eq:heat_flux_vec}
    \Vec{Q} = n\left\langle \frac{1}{2} m U^2 \Vec{U} \right\rangle
    \quad
    \text{and}
    \quad
    \Vec{Q}_\omega = n \left\langle \frac{1}{2} I W^2 \Vec{U} \right\rangle
\end{equation}
where
\begin{equation}
    U_i = c_i - u_i
    \quad
    \text{and}
    \quad
    W = \omega - \Bar{\omega}.
\end{equation}
Table \ref{hydro_variables} gives a summary of these hydrodynamic quantities and the corresponding fluxes.
By construction, the pressure tensor $P_{ij} = P_{ji}$ is symmetric at this level of description.
The equations of change \eqref{equations_of_change} are exact but not closed, because the values of the fluxes are not yet determined as a function of the hydrodynamic variables.

\begin{table}[]
\centering
\begin{tabular}{lll}
\toprule
quantity & density & current \\
\midrule
mass                    & $\rho$          & $\rho u_i$ \\
linear momentum         & $u_i$           & $P_{ij}$ \\
angular momentum        & $\bar{\omega}$  & $L_i$ \\
translational energy    & $e_t$           & $Q_i$ \\
rotational energy       & $e_r$           & $Q_{\omega,i}$ \\
\bottomrule
\end{tabular}
\caption{Hydrodynamic variables and their currents}
\label{hydro_variables}
\end{table}

Note that the linear momentum field $\rho \vec{u}$ we define is obtained from the center-of-mass momentum $m\vec{c}$, which is different from the \enquote{total momentum} considered in Ref.~\cite{Markovich2024}
(see also Refs.~\cite{rosenfeld1940energy,Belinfante1940,markovich2021odd, markovich2022nonreciprocity,Bliokh2025,Morrison2014}), which can be obtained as a linear combination of $\rho \vec{u}$ and $\bar{\omega}$, but is not necessarily conserved.

\subsection{Hydrodynamic Currents}
\label{transport_coefficients}





\subsubsection{Linear response}

In order to close the equations of change obtained in Sec.~\ref{equations_of_change} into hydrodynamic equations, we must express the pressure tensors and heat flux vectors as a function of the hydrodynamic variables.
To do so, we assume linear response and express the fluxes as a linear function of the local variations in the hydrodynamic quantities (relaxation currents) and their gradients (transport terms). 

Generally, a variation in any hydrodynamic variable $\delta \rho_\beta$ can generate a current $J_{\alpha,i}$ associated with any hydrodynamic variable $\rho_\alpha$, that is,
\begin{equation}
    \label{eq:general_linear_response_currents}
    J_{\alpha,i} = J^{(rel)}_{\alpha,i} + J^{(TP)}_{\alpha,i} \equiv \zeta^i_{\alpha\beta} \delta \rho_\beta + \Lambda^{ij}_{\alpha \beta} \partial_j \delta \rho_\beta,
\end{equation}
where sum over repeated indices is implied and $\zeta^i_{\alpha\beta}$ and $\Lambda^{ij}_{\alpha \beta}$ are coefficients, respectively called relaxation coefficients ($\zeta^i_{\alpha\beta}$) and transport coefficients ($\Lambda^{ij}_{\alpha \beta}$) that depend on the microscopic details of the system.
These coefficients contain all the familiar quantities such as heat conductivity or viscosity, arranged in a way convenient for calculations.
Kinetic theory gives a way to calculate these coefficients, and their values for our microscopic model are one of the main results of this paper, details of which are given in Sec. \ref{sec:Adiabatic_Elimination}.

For a conventional fluid, the viscous term (momentum current due to momentum gradients) $ -\eta_{ijk\ell} \partial_\ell v_k $ and the heat conduction term (temperature current due to temperature gradients) $-\lambda_{ij} \partial_{j} T$ are examples of transport currents, and the viscosity tensor $\eta_{ijkl}$ and the heat conductivity $\lambda_{ij}$ are the associated transport coefficients. On the other hand, the linear perturbation on hydrostatic pressure $J^{(rel)}_{v_i,j} = \delta_{i,j}(\delta n T_t + n \delta T_t)$ is an example of a relaxation current.

\subsubsection{Linearized hydrodynamic equations}

We first linearize the equations of change \eqref{equations_of_change} by considering small perturbations
$\rho_\alpha \equiv \rho_\alpha^{(0)} + \delta \rho _\alpha$ from the equilibrium values $\rho^{(0)}_\alpha$.
At lowest order, we find
\begin{subequations}
\label{eq:linear_hydro}
\begin{align}
    D_t \delta \rho + \rho^{(0)} \nabla \cdot \Vec{\delta u} = 0\\
    \rho^{(0)} D_t \delta u_i + \partial_j P_{ij}^{(1)} = 0\\
    I n^{(0)} D_t \delta \omega + \partial_i M_i^{(1)} - I d \delta n = \mathcal{C}^{(1)}[I \omega]\\
    n^{(0)} D_t \delta e_t + \nabla \cdot \Vec{Q}^{(1)} + P_{ij}^{(0)} \partial_i \delta u_j = \mathcal{C}^{(1)}[mv^2/2]\\
    \label{eq:linear_hydro2}
    n D_t \delta e_r + \nabla \cdot \Vec{Q}_\omega^{(1)} = \mathcal{C}^{(1)}[I\omega^2/2 - I\omega \Bar{\omega}],
\end{align}
\end{subequations}
where, $\mathcal{C}^{(1)}[\psi]$ are the first order terms in the integral $\mathcal{C}[\psi]$.
Using the definitions (\ref{eq:pressure_tensor}-\ref{eq:heat_flux_vec}), we find
\begin{align}
    M^{(0)}_i = 0
    \qquad
    \Vec{Q}^{(0)} = 0
    \qquad
    \Vec{Q}^{(0)}_\omega = 0
\end{align}
as well as
\begin{align}
    P_{ij}^{(0)} = n^{(0)} k_B T_t^{(0)}.
\end{align}
All the first order contributions $P_{ij}^{(1)}, M_i^{(1)} Q^{(1)}_i$ and $Q_{\omega,i}^{(1)}$ correspond to the linear response terms are introduced in equation \eqref{eq:general_linear_response_currents} and can be numerically calculated using the method of adiabatic elimination, details of which are given in the following section.

\subsubsection{Adiabatic Elimination}
\label{sec:Adiabatic_Elimination}

In order to make connection with the microscopic description of the system, we use the adiabatic elimination method \cite{dorfman2021contemporary,Resibois1970}
which is equivalent to the usual Chapman-Enskog theory \cite{chapman1990mathematical}. 
The general idea of adiabatic elimination is to split the Boltzmann equation into slow and fast variable subspaces, assume that fast variables relax very quickly to equilibrium and do not change in time, and obtain them in terms of slow variables. Consequently, we obtain the linear hydrodynamic equations for the slow variables that effectively contain the information of the fast variables.

Projection into slow and fast spaces is done via the inner product defined in Eq. \eqref{eq:inner-product}, and
to execute further steps, we first need to construct an orthonormal basis of functions $\Phi_\alpha$ for the slow variable space that satisfies $(\Phi_\alpha,\Phi_\beta) = \delta_{\alpha \beta}$, with respect to this inner product. As we mentioned in Sec. \ref{hydrodynamic_variables} we will choose the hydrodynamic variables to be the density $\rho$, linear momenta $\rho\vec{u}$, angular momenta $In\omega$, and translational $T_t$ and rotational temperatures $T_r$.
In 2D, the orthonormal basis that corresponds to these variables is
\begin{equation}
    \begin{pmatrix}
        \Phi_m\\
        \Phi_{v,i}\\
        \Phi_{\omega}\\
        \Phi_{e,t}\\
        \Phi_{e,r}
    \end{pmatrix} =     \begin{pmatrix}
        1\\
        \sqrt{\beta_t m} v_i\\
        \sqrt{\beta_r I}\omega\\
        \left[\beta_t m \frac{c^2}{2} - 1 \right]\\
        \sqrt{2}\left[ \beta_r I \frac{\omega^2}{2} - \frac{1}{2} \right]
    \end{pmatrix},
\end{equation}
which leads to the corresponding linear perturbations of hydrodynamic functions  as
\begin{subequations}
\label{eq:perturbation_as_inner_prod}
\begin{align}
    (\Phi_m,\phi) &= \frac{\delta \rho}{\rho^{(0)}}\\
    (\Phi_{v,i},\phi) &= \sqrt{\beta_t^{(0)} m} \delta u_i = u_i - u^{(0)}_i\\
    (\Phi_{\omega},\phi) &= \sqrt{\beta_r^{(0)} I} \delta \bar{\omega
    } = \bar{\omega} - \bar{\omega
    }^{(0)}\\
    (\Phi_{e,t},\phi) &= \frac{\delta T_t}{T_t^{(0)}}\\
    (\Phi_{e,r},\phi) &= \frac{\delta T_r}{\sqrt{2} T_r^{(0)}}.
\end{align}
\end{subequations}

Now, with the orthonormal basis we constructed, we can define a projection operator $P$ to the slow variable space as,
\begin{equation}
    P = \sum_{\alpha \in \text{slow}} |\Phi_\alpha)(\Phi_\alpha|,
\end{equation}
which is a Hermitian operator, i.e., $P^\dagger = P$. Conversely, the projection operator $Q$ to the fast variable space becomes,
\begin{equation}
    Q = \mathbb{I} - P = \mathbb{I} - \sum_{\alpha \in \text{slow}} |\Phi_\alpha ) ( \Phi_\alpha|,
\end{equation}
which also satisfies $Q^\dagger = Q$.
Any function $\phi$ can be projected onto the slow or fast variable space through
\begin{align}
      \phi_S &\equiv \sum_{\alpha \in \text{slow}} (\phi, \Phi_\alpha) |\Phi_\alpha)\\
     \phi_F &\equiv \phi - \sum_{\alpha \in \text{slow}} (\phi, \Phi_\alpha) |\Phi_\alpha).     
\end{align}
Similarly, any operator acting on this function space can be decomposed into a block form as
\begin{equation}
    L = \begin{pmatrix}
        L_{SS} & L_{SF}\\
        L_{FS} & L_{FF}
    \end{pmatrix},
\end{equation}
where $L_{SS} = PLP$, $L_{SF} = PLQ$, $L_{FS} = QLP$ and $L_{FF} = QLQ$. 

Then using the projection operators we defined, the linear Bolzmann equation in Fourier space \eqref{eq:lin_Boltzmann_eq} can be rewritten as,
\begin{equation}
    \partial_t \begin{pmatrix}
        \phi_S \\
        \phi_F
    \end{pmatrix} = L(\Vec{q}) \begin{pmatrix}
                \phi_S \\
        \phi_F
    \end{pmatrix},
\end{equation}
where,
\begin{equation}
    L(\Vec{q}) = \begin{pmatrix}
        -iq_i c^i_{SS} + L_{SS} & -iq_i c^i_{SF} + L_{SF}\\
         -iq_i c^i_{FS} +L_{FS} & L_{FF}
    \end{pmatrix}.
\end{equation}
Then, by assuming that the fast variables relax very quickly to their equilibrium values, i.e., $\partial_t \phi_F \approx 0$ we obtain an expression for $\phi_F$ in terms of $\phi_S$ in the first equation. This effectively integrates out the fast variables, and yields a linear equation only in terms of the slow variables. First setting $\partial_t \phi_F$ gives,
\begin{equation}
    \phi_F \approx L_{FF}^{-1}(iq_i c^i_{FS} - L_{FS}) \phi_S,
\end{equation}
and inserting this back to the first equation gives,
\begin{equation}
    \partial_t \phi_S = L_{\text{eff}} \phi_S,
\end{equation}
where,
\begin{align}
    L_{\text{eff}} &= (-i q_i c^i_{SS} + L_{SS} ) \nonumber \\ &+ (-i q_i v^i_{SF} + L_{SF} ) \left( L_{FF}' \right)^{-1}(-i q_i c^i_{FS} + L_{FS} ).
    \label{eq:effective_collision_operator}
\end{align}

From this expression, the linearized hydrodynamic equations can be obtained by taking the inner product of both sides with slow variables $\Phi_\alpha$. 
This yields
\begin{equation}
    \partial_t \label{eq:Boltzmann-hydro}
    (\Phi_\alpha , \phi) = \left[ L_{\text{eff}} \right]_{\alpha \beta} (\Phi_\beta,\phi),
\end{equation}
where the matrix elements are defined as 
\begin{equation}
    \left[L_{\text{eff}} \right]_{\alpha \beta} = \left( \Phi_\alpha, L_{\text{eff}}\Phi_\beta \right).
\end{equation}
Comparing these equations with Eqs.~\eqref{eq:general_linear_response_currents} and~\eqref{eq:linear_hydro} allows us to identify explicit expressions for the transport and relaxation coefficients (as well as the collision terms and the external torque term) as an inner product of the slow space vectors. Particularly, a direct comparison shows that the transport coefficients are
\begin{equation}
    \label{eq:transport_coef_inner_product}
    \Lambda^{\alpha\beta}_{ij} = (Qc_i\Phi_\beta,L^{-1}_{FF}c_j \Phi_\alpha)
\end{equation}
and the relaxation coefficients are
\begin{align}
    \label{eq:relaxation_current_inner_product}
    \zeta^i_{\alpha\beta} &= (\Phi_\alpha,c^i_{SF}L^{-1}_{FF} L_{FS} \Phi_\beta) \\ &+ (\Phi_\alpha,L_{SF}L^{-1}_{FF}c^i_{SF}\Phi_\beta) + (\Phi_\alpha, c^i_{SS}\Phi_\beta).
\end{align}
The viscosity tensor $\eta_{ijk\ell} = \Lambda^{v_kv_\ell}_{ij}$ can be obtained by choosing $\Phi_\alpha = \Phi_{v,k}$ and $ \Phi_\beta = \Phi_{v,\ell}$, while the heat conductivity tensor $\lambda_{ij} = \Lambda^{e_te_t}_{ij}$ is obtained by choosing $\Phi_\alpha = \Phi_\beta = \Phi_{e,t}$. An explicit version of \eqref{eq:Boltzmann-hydro} is given in App. \ref{app:full_hydro_adiabatic}, with color codes identifying different types of terms, including all the other relaxation and transport terms, as well as the external torque term and the collision terms for non-conserved variables.\footnote{Note that if the current terms are written strictly in the form of \eqref{eq:general_linear_response_currents}, some coefficients will be multiplied by the equilibrium values of the hydrodynamic variables, for instance $\Lambda^{e_t e_r}_{ij} \rightarrow \sqrt{\frac{T_tI}{T_r}}\Lambda^{e_t e_r}_{ij}$, which will depend on the microscopic details of the system. All the terms are explicitly shown in App. \ref{app:full_hydro_adiabatic}.   }  


\subsubsection{Nondimensionalization}
To obtain the full linear description of our model, we need to compute the inner products given in Eqs. \eqref{eq:transport_coef_inner_product} and \eqref{eq:relaxation_current_inner_product}. It is convenient to introduce dimensionless quantities to compute these inner products numerically. We present our procedure below for 2D rotating disks, but it is straightforward to generalize to systems with different degrees of freedom. Note also that while this procedure has the same motivation as what we do in the stability analysis (App. \ref{app: Stability Analysis}), the steps are slightly different.

We first introduce the dimensionless velocities,
\begin{equation}
    \tilde{c}_i = \sqrt{\frac{\beta_t m}{2}} c_i, \tilde{\omega} = \sqrt{\frac{\beta_r I}{2}} \omega,
\end{equation}
which leads to the dimensionless distribution function $\tilde{f}$ via,
\begin{equation}
    \tilde{f} \equiv  \frac{1}{n} {\frac{2\pi T_t}{m}}\sqrt{\frac{2\pi T_r}{I}} f,
\end{equation}
which is simply the coefficient in front of the equilibrium Gaussian distribution, but dimension-wise, it can be used for any distribution function.
We can furthermore define the dimensionless version of the inner product (say of two dimensionless quantities $\phi$ and $\chi$) as,
\begin{equation}
    \label{eq:non-dim_inner_prod}
    (\phi,\chi) = \frac{1}{\pi^{3/2}}\int d\vec{\tilde{c}} d\tilde{\omega} \tilde{f} \Bar{\phi}\chi = \frac{1}{\pi^{3/2}} (\phi,\chi)_{ND},
\end{equation}
and the dimensionless linear collision operator as,
\begin{equation}
    \frac{4\sqrt{2} a n}{\sqrt{\beta_t m}\pi^{3/2}} \frac{1}{2} \int d\tilde{\vec{c}}_{2} d\tilde{\omega}_2 d\apsevector |\tilde{\vec{c}}_{12}\cdot\apsevector| \tilde{f} \Delta \phi = \frac{4\sqrt{2} a n}{\sqrt{\beta_t m}\pi^{3/2}} \tilde{C}[\phi].
\end{equation}
This definition of the dimensionless collision operator, also dictates how the Boltzmann operator $L \equiv C[\phi] - d\partial_\omega[\phi] - d \partial_\omega\left(\log f^{(0)}\right)[\phi]$ is non-dimensionalized, which is,
\begin{align}
    L[\phi] &= C[\phi] - d\partial_\omega \phi - d \partial_\omega \left(\log f^{(0)}\right) \phi  \nonumber \\ 
    &= \frac{4\sqrt{2} a n }{\sqrt{\beta_t m} \pi^{3/2}} \left[ \tilde{C}[\phi] -  \frac{\text{sgn}(d)}{\sqrt{\tilde{T}_t} \tilde{T}_r } \frac{\sqrt{q} \pi^2 }{\tilde{G}} \tilde{L}_\omega \right] \\
    &\equiv \frac{4\sqrt{2} a n }{\sqrt{\beta_t m} \pi^{3/2}} \tilde{L}[\phi],
\end{align}
where non-dimensional temperatures are defined as $\tilde{T}_{t,r} = \frac{T_{t,r}}{ma^2|d|}$ as in App. \ref{app:dimless_quant} and we introduced the shorthand notation $\tilde{L}_\omega \equiv \partial_{\tilde{\omega}}[\phi] + \partial_{\tilde{\omega}}(\log \tilde{f}^{(0)})  $.
Since $\Phi_\alpha$ are already dimensionless, this is sufficient to obtain the dimensionless transport coefficients $\tilde{\Lambda}^{\alpha \beta}_{ij}$ as
\begin{align}
    \label{eq:transport_coeff}
    &(Q c_i \Phi_\beta, \mathcal{C}^{-1}_{FF} Q c_j \Phi_\ell ) = \nonumber \\
    &\frac{1}{2\sqrt{2}\sqrt
    {\beta_t m} a n} ( Q \tilde{c}_i \Phi_\beta, \tilde{\mathcal{C}}_{FF}^{-1} Q\tilde{c}_j \Phi_\ell )_{ND} \equiv \nonumber \\
    & -\frac{1}{2\sqrt{2}\sqrt
    {\beta_t m} a n} \tilde{\Lambda}^{\alpha \beta}_{ij},
\end{align}
constant factors may be absorbed into it when $\Phi_\alpha$ are written in terms of conventional hydrodynamic functions. The dimensionless relaxation current $\tilde{\zeta}^i_{\alpha,\beta}$ can also be found using a similar procedure, which gives,
\begin{align}
    \label{eq:relaxation_coef}
    \zeta^i_{\alpha\beta} &= \sqrt{\frac{2}{\beta_t m}}\Big[ (\Phi_\alpha, \tilde{c}^i_{SF}\tilde{L}^{-1}_{FF} \tilde{L}_{FS} \Phi_\beta )\\ \nonumber
    &+ (\Phi_\alpha,  \tilde{L}_{SF} \tilde{L}^{-1}_{FF} \tilde{c}^i_{SF} \Phi_\beta )+ (\Phi_\alpha, \tilde{c}^i_{SS}\Phi_\beta) \Big] \\ \nonumber
    &\equiv \sqrt{\frac{2}{\beta_t m}} \tilde{\zeta}^i_{\alpha,\beta}.
\end{align}

\begin{figure}[b]
\includegraphics[scale=0.35]{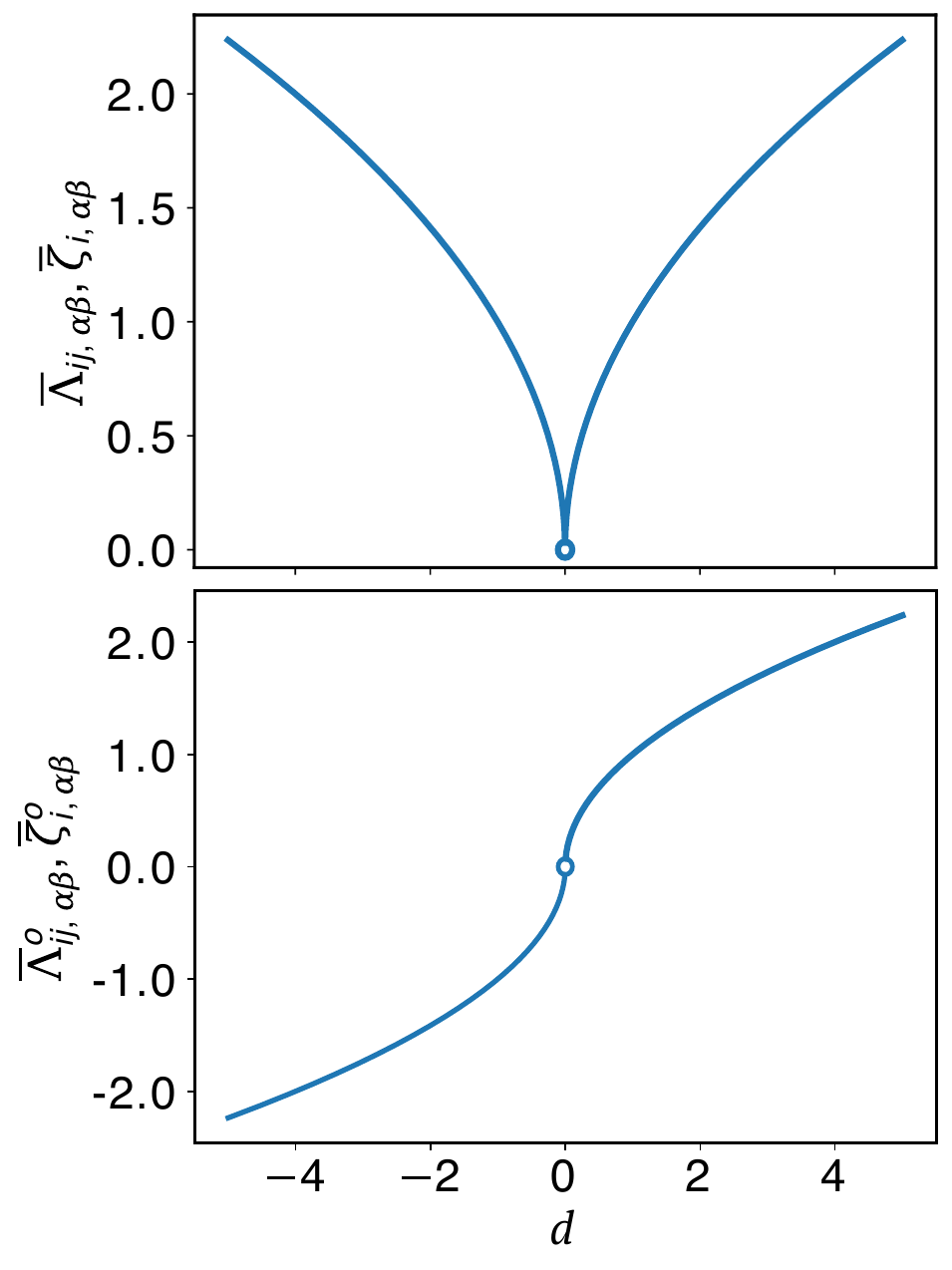}
\caption{ \textbf{Torque Dependence of Response Coefficients:} The square root dependence of response coefficients is depicted. Plotted values $\overline{\Lambda}_{ij,\alpha\beta}(d) \equiv \Lambda^{\alpha\beta}_{ij}(d)/\Lambda^{\alpha\beta}_{ij}(d=1)$ are the ratio of the coefficients divided by a reference value evaluated at $d=1$. The upper plot represents the response coefficients that are even functions of the field $\Lambda^{\alpha\beta}_{ij}(-d) = \Lambda^{\alpha\beta}_{ij}(d)$ and the lower plot represents the ones that are odd functions of the field $\Lambda^{o}_{ij,\alpha\beta}(-d) = -\Lambda^{o}_{ij,\alpha\beta}(d)$. }
\label{fig:torque_dependence}
\end{figure}

Note that for both terms, the only dependence on the external field in comes from the factor
\begin{equation}
    1/\sqrt{\beta_t} \sim \sqrt{|d|}
\end{equation}
which makes the torque dependence of all transport coefficients explicit, as represented in Figure \ref{fig:torque_dependence}. 
We defined the overline response coefficients as the ratio of the coefficient and a reference value evaluated at unit torque, that is
\begin{equation}
    \overline{\Lambda_{ij,\alpha\beta}} \equiv \frac{\Lambda_{ij}^{\alpha\beta}(d)  }{ \Lambda_{ij}^{\alpha\beta}(d=1) }
\end{equation}
and the superscript $o$ is used to separate the part of the tensor that is odd under the external field $d$.

\section{Numerical Computation of the Relaxation and Transport Coefficients}
\label{numerical_computation}



The computational expressions for the relaxation and transport coefficients can be obtained by inserting the corresponding basis vectors into the non-dimensional inner products \eqref{eq:relaxation_coef} and, \eqref{eq:transport_coeff}, respectively. The expressions for the transport coefficients in the irreducible representation are listed in Eqs. \eqref{eq:all_relaxation_coef} and \eqref{eq:all_transport_coef} in terms of these inner products. 

\begin{figure*}
\includegraphics[scale=0.45]{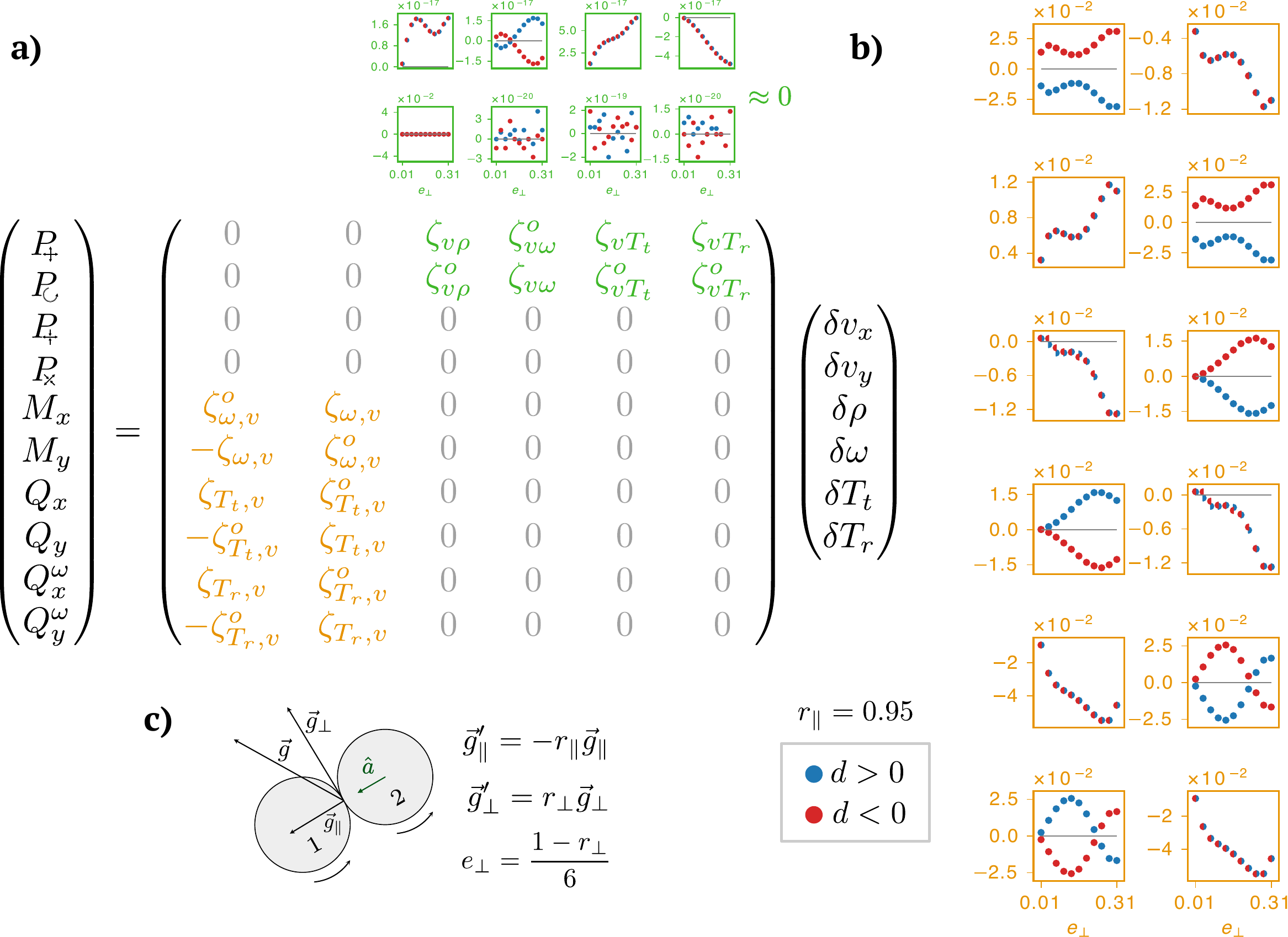}
\caption{\label{fig:relaxation_matrix_irrep} \textbf{Relaxation coefficients. a)} The matrix equation describing the relaxation coefficients is given in the irreducible representation (App. \ref{app:Irrep}), for a rotationally invariant fluid. 
The pressure tensor is split into independent components of volume expansion, rotation and shear rates. In the irreducible representation, relaxation coefficients are split into block matrices, off diagonal elements of which are associated with transverse responses. Coefficients that are odd in the external field denoted by the superscript $o$. As can be seen, for the coefficient matrices that couple the angular momentum to the other hydrodynamic fields, diagonal entries are odd, while the off-diagonal entries are odd for all the other coefficient matrices. Although the top right section of the matrix colored in green is allowed by the symmetry, numerical calculations show that they are orders of magnitude smaller than the rest of the terms, except the perturbations to the hydrostatic pressure, which can exactly be calculated.  \textbf{b)}The computed values of the relaxation coefficients on the bottom left of the matrix are depicted, with the same color. The upper right block turns out to be zero after explicit computation except the perturbations to the hydrostatic pressure, which can be exactly calculated. $x=0$ axis is highlighted with a thin gray line where it is relevant. As can be seen some coefficients cross this line, showing that the coefficients can change sign without the torque changing direction. \textbf{c)} A graphical depiction of the restitution coefficients $\restcoefnormal$, $\restcoeftangent$ and $\modtangentfric$. After the collisions the velocity of the point of contact parallel to the apse vector $\vec{g}_\parallel$ is reduced by the factor $\restcoefnormal$ and the part tangent to the apse vector $\vec{g}_\bot$ is reduced by the factor $\restcoeftangent$. $\modtangentfric$ depends on $\restcoeftangent$ and is more convenient to work with in computations. For 2D hard disks $\modtangentfric = (1-\restcoeftangent)/6$. }
\end{figure*}

\begin{figure*}
\includegraphics[scale=0.54]{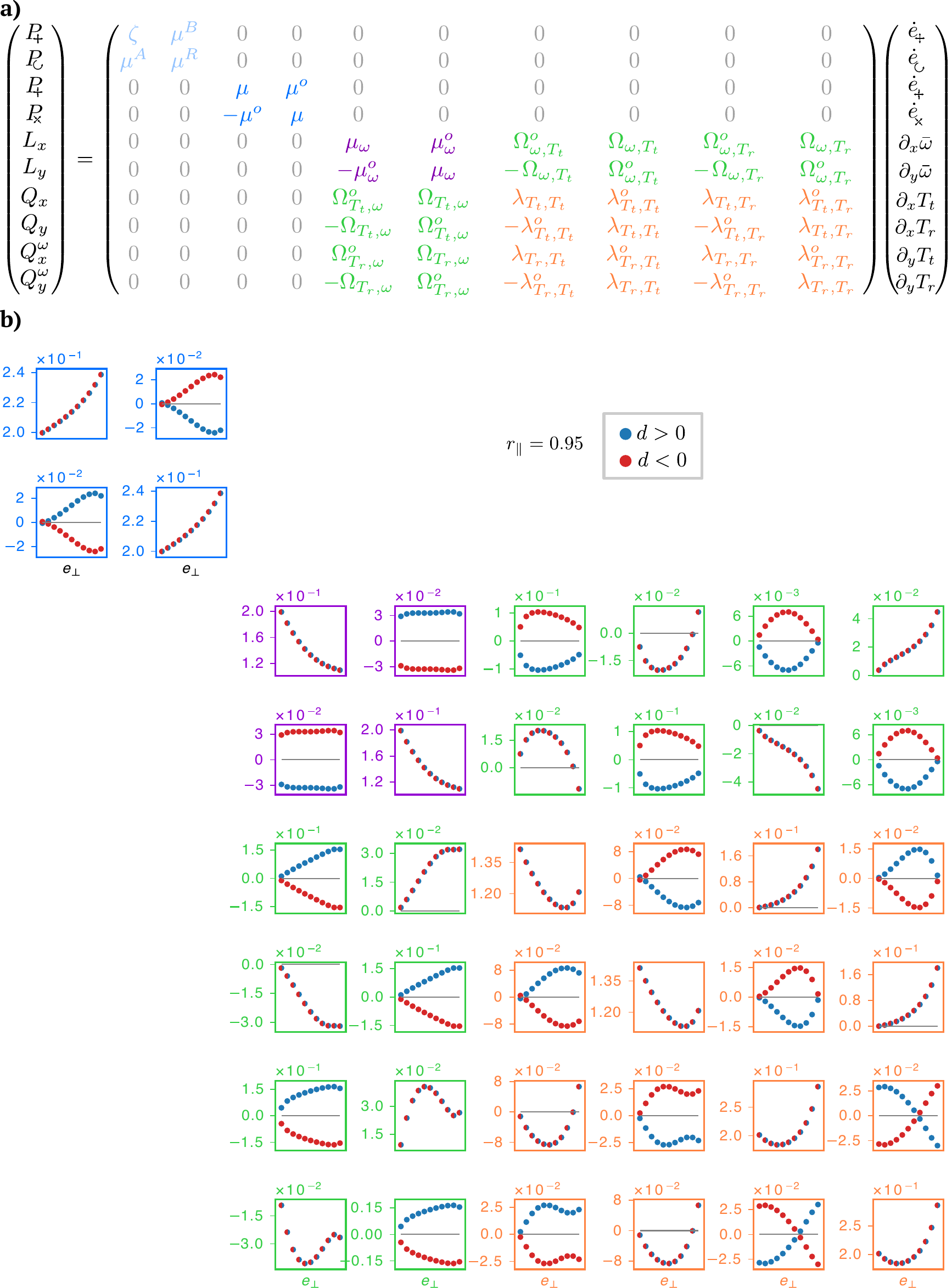}
\caption{\label{fig:transport_matrix_irrep} \textbf{Transport coefficients.} The matrix equation describing the transport coefficients is given in the irreducible representation (App. \ref{app:Irrep}) for a rotationally invariant fluid. 
The pressure and the deformation rate tensors are split into independent components of volume expansion, rotation and shear rates. Moreover, in the irreducible representation, transport tensors split into block matrices with anti-symmetric off-diagonal components, which are associated with the transverse responses. Coefficients that are odd in the external field are denoted by the superscript $o$. For the coefficient matrices that couple angular momentum with the other hydrodynamic variables, the diagonal elements are odd, while the off-diagonal elements are odd for all the other coefficient matrices. \textbf{b)}The computed values of the transport coefficients are depicted, with the same colors. The upper left block is zero due to collisions being instantaneous. $x=0$ axis is highlighted with a thin gray line where it is relevant. As can be seen some coefficients cross this line, showing that the coefficients can change sign without the torque changing direction.}
\end{figure*}

In a system with rotational symmetry, expressions for relaxation and transport currents are given in Eq. \eqref{eq:general_linear_response_currents} and will be restricted to the form shown in Figures \ref{fig:relaxation_matrix_irrep}.a and \ref{fig:transport_matrix_irrep}.a, respectively. The matrices presented in the figures are written in the irreducible representation explained in App. \ref{app:Irrep}. Coefficients that are odd in the external field $d$ are denoted with the superscript $o$.

In our model, explicit numerical computation shows that all relaxation coefficients $\zeta_{v \beta}, \zeta^o_{v\beta}$ that couple variations in scalar fields $\delta \rho_\beta$ to the pressure are zero except the perturbations to the hydrostatic pressure $\zeta_{v\rho} = 1/(\rho \beta_t m)$, $\zeta_{v T_t} = 1/m$, and the rest of the coefficients are non-zero. For the transport coefficients, viscosities that resist volume expansion, rotations and those that couple these two to each other are $\zeta = \eta^B = \eta^A = \eta^R = 0$, due to having instantaneous collisions, while all the other ones are non-zero. We present the numerical values of all non-zero relaxation and transport coefficients in Figures 
\ref{fig:relaxation_matrix_irrep}.b and \ref{fig:transport_matrix_irrep}.b, respectively. The colors used to group the coefficients in the matrix representation are the same for the plots with computed values. We choose a fixed normal restitution $\restcoefnormal$ and plot their values over a range of tangential restitution $\modtangentfric$. 
Perhaps surprisingly, we find that odd coefficients can change sign as the tangential restitution is varied, even though the torque direction remains the same. 



\section{Conclusion}
We analyzed a minimal but realistic collision-based model for odd fluids using the kinetic theory of granular gases.
We showed that transport and relaxation coefficients have components that are odd in the driving field, a hallmark of odd fluids.
We also showed that all of these coefficients are proportional to the square root of the driving field, and have analyzed their behavior across different values of friction parameters by numerical computation. 
Our results indicate that 
the macroscopic response of the fluid is the result of a complex interplay between the drive and the collisions, as manifested by features such as the change of sign of some response coefficients without changing the direction of the driving field.
We furthermore discuss the choice of hydrodynamic variables of the theory by analyzing the eigenvalue spectrum of the collision operator.
Our methodology 
can be applied to other collision models and to systems in higher dimensions with minimal modification.

\begin{acknowledgements}
We thank Daniel Seara for feedback on the manuscript.
V.V. acknowledges partial support from the Army Research Office under grant W911NF-22-2-0109 and W911NF-23-1-0212.
M.F. and V.V acknowledge partial support from the France Chicago center through a FACCTS grant.
This research was partly supported from the National Science Foundation through the Center for Living Systems (grant no. 2317138), the National Institute for Theory and Mathematics in Biology, the Simons Foundation and the Chan Zuckerberg Foundation. This work was completed in part with resources provided by the University of Chicago’s Research Computing Center.
\end{acknowledgements}


\clearpage

 \appendix
\section{Derivation of the Collision Term}
The arguments that lead to Eq. \eqref{eq:collision_operator}closely follow the treatment in Ref. \cite{dorfman2021contemporary} chapter 2. We will split the collision operator into two parts $\fullcollop = \Gamma^+ - \Gamma^-$, the first denoting the gain term and the second denoting the loss term. In the sections below, we derive each term from the collision rules assuming Boltzmann's molecular chaos assumption. 
\begin{figure}
\includegraphics{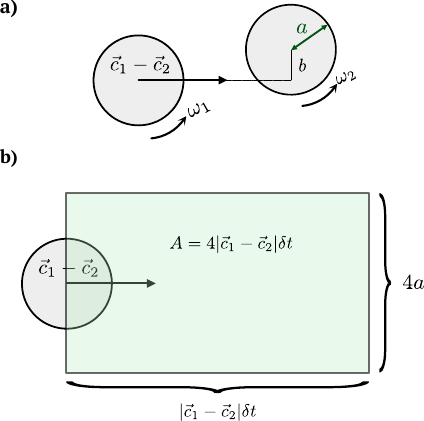}
\caption{ \textbf{Microscopic Collisions and the Impact Factor:} \textbf{a)} Depiction of the collision in the center of mass frame of the second particle The frame is rotated so that the the velocity of the first particle $\vec{c}_1 - \vec{c}_2$ is parallel to the x-axis. The impact factor $b$ is the vertical distance between the centers of mass of the particles in this frame. \textbf{b)} Depiction of the collision cylinder formed within a time frame $\delta t$ by a particle of radius $a$ moving with velocity $\vec{c}_1 - \vec{c}_2$.}
\label{fig:pair_collision_appendix}
\end{figure}
\subsection{Derivation of the Loss Term $\Gamma^-$}

Let us consider two colliding hard disks with radius $a$, one with linear and angular velocities $(\vec{c}_1,\omega_1)$ and the second one with $(\vec{c}_2,\omega_2)$, and suppose that we move to the coordinate frame of the center of mass of the second particle. In this frame, the center of mass of the first particle will move with $\vec{c}_1 - \vec{c}_2$. If we then rotate the axis so that the velocity $\vec{c}_1 - \vec{c}_2$ is aligned with the x-axis (which we can always do without a loss of generality), then the collision in 2D can be completely characterized by the velocities $\vec{c}_1,\vec{c}_2,\omega_1,\omega_2$ and the impact factor $b$, which denotes the vertical distance between the centers of mass between the two disks. Note that for the derivation of the collision operator, considering the center of mass velocities as we change the reference frame will be sufficient. The depiction of the collision in the mentioned reference frame is given in Fig. \ref{fig:pair_collision_appendix}.a.

It is easy to see that a collision can happen only if $b < 2a$. Let us next define the collision cylinder for particle 1 in the reference frame of particle 2 (which in 2D is a rectangle) as the area covered within a time frame $\delta t$ where a collision will happen if another particle comes in. The area $A$ of this cylinder is,
\begin{equation}
    A = 4a \delta t |\vec{c}_2 - \vec{c}_1|.
\end{equation}
A pictorial representation of the collision cylinder is given in Fig. \ref{fig:pair_collision_appendix}.b.
If the particle 2 is in this cylinder, then a collision will happen within a time-frame $\delta t$, assuming that no other particle will intervene. We also assume that no other collision is on-going at time $t$ to keep everything simple.

\subsubsection{Molecular Chaos Assumption (Stosszahlansatz)}
To write down the expression for the loss term, we follow Boltzmann's argument, which proceeds as follows: If the single particle distribution function is $f(\vec{r},\vec{c},\vec{\omega},t)$, 
\begin{itemize}
    \item At a given time $t$, the number of particles with velocity $\vec{c}_1,\Vec{\omega}_1$ in the phase-space volume $\delta \vec{c}_1 \delta \Vec{\omega}_1\delta \Vec{r}_1$ is $f(\Vec{r}_1,\vec{c}_1,\Vec{\omega}_1)\delta \vec{c}_1 \delta \Vec{\omega}_1\delta \Vec{r}_1$.
    \item There is a collision cylinder attached to each particle and the total area of these collision cylinders is $ 4a  |\vec{c}_2 - \vec{c}_1| f(\Vec{r}_1,\vec{c}_1,\Vec{\omega}_1)\delta \vec{c}_1 \delta \Vec{\omega}_1\delta \Vec{r}_1 \delta t$.
    \item Since it is impossible to track the dynamical history of each particle, we will proceed with the statistical argument of Boltzmann: Every particle 2 that lies inside the collision cylinder will cause a collision and the number of these collisions inside the cylinders will be determined probabilistically by $f(\Vec{r}_1,\vec{c}_2,\Vec{\omega}_2) \delta \vec{c}_2 \delta\Vec{\omega}_2 \times 4a  |\vec{c}_2 - \vec{c}_1| f(\Vec{r}_1,\vec{c}_1,\Vec{\omega}_1)\delta \vec{c}_1 \delta \Vec{\omega}_1\delta \Vec{r}_1 \delta t$. The first term is the number of particles with velocities $\vec{c}_2,\Vec{\omega}_2$ per unit area at $\Vec{r}_1,t$ and the second term is the total area of the collision cylinders.
    \item Then the loss term $\Gamma^-$ is simply obtained by integrating this term over all possible values $\vec{c}_2, \Vec{\omega}_2$, since any collision with any particle 2 will change the velocity $\vec{c}_1,\Vec{\omega}_1$ of particle 1 resulting in a loss.
\end{itemize}
Applying the full argument, we have,
\begin{align}
     \Gamma^- \delta \vec{c}_1 d\Vec{\omega}_1 \delta \Vec{r}_1 \delta t = 4a\int d\vec{c}_2 d\Vec{\omega}_2 |\vec{c}_2 - \vec{c}_1| \nonumber \\ \times f(\Vec{r}_1,\vec{c}_1,\Vec{\omega}_1,t)f(\Vec{r}_1,\vec{c}_2,\Vec{\omega}_2,t) \delta \vec{c}_1 \delta \Vec{\omega}_1 \delta \Vec{r}_1 \delta t,   
\end{align}

which implies,
\begin{equation}
    \Gamma^-  = 4a\int d\vec{c}_2 d\Vec{\omega}_2 |\vec{c}_2 - \vec{c}_1| f(\Vec{r}_1,\vec{c}_1,\Vec{\omega}_1,t)f(\Vec{r}_1,\vec{c}_2,\Vec{\omega}_2,t).
\end{equation}

\subsection{Derivation of the Gain Term $\Gamma^+$}
We will follow a similar argument as we did in the previous section. However, this time, the task at hand is a bit more difficult as we need to find the pair of velocity sets $(\vec{c}_1'',\Vec{\omega}_1''), (\vec{c}_2'',\Vec{\omega}_2'')$ that will yield velocities $(\vec{c}_1,\omega_1)$.

Let us construct the collision cylinders again. This time, however, since the outcome of the collisions, hence the pre-collision velocities, depend on the impact parameter itself, we need to start with the infinitesimal collision cylinders, height of which is $db$. Then, we can apply the molecular chaos assumption as before, which follows as:
\begin{itemize}
    \item The number of particles with velocity $\vec{c}_1'',\Vec{\omega}_1''$ at $\Vec{r}_1$ within a phase space volume of $\delta \vec{c}_1'' \delta \Vec{\omega}_1'' \delta \Vec{r}_1$ at a time $t$ is $f(\Vec{r}_1,\vec{c}_1'',\Vec{\omega}''_1,t)\delta \vec{c}_1'' \delta \Vec{\omega}_1'' \delta \Vec{r}_1$.
    \item The area of each infinitesimal collision cylinder of height $db$ is $db |\Vec{c}''_{12}|\delta t$ and the total area of the collision cylinders is $f(\Vec{r}_1,\vec{c}_1'',\Vec{\omega}_1'',t) |\Vec{c}''_{12}|\delta t \delta \vec{c}_1'' \delta \Vec{\omega}_1'' \delta \Vec{r}_1 db$.
    \item The total number of collisions is equal to the number of collision cylinders and the number of particles with velocities $(\vec{c}_2'',\Vec{\omega}_2'')$ per unit area at around $\Vec{r}_1$. This means, the total number of collisions become, $f(\Vec{r}_1,\vec{c}_1'',\Vec{\omega}_1'',t)f(\Vec{r}_1,\vec{c}_2'',\Vec{\omega}_2'',t) 
 |\Vec{c}_{12}''| \times $
 $\delta t \delta \vec{c}_1'' \delta \Vec{\omega}_1'' \delta \vec{c}_2'' \delta \Vec{\omega}_2'' \delta \Vec{r}_1 db$.
\end{itemize}
\subsubsection{The Apse Vector $\apsevector$}
It is difficult to perform the calculations in terms of the impact factor $b$ and it is much more convenient to describe the collisions with the apse vector $\apsevector$ which we defined in the main text. In 2D, this vector can be parameterized by an angle $\phi$, such that $\apsevector = (\cos\phi, \sin \phi)$. Then, the impact parameter $b$ can be expressed, in terms of this angle as $b = 2a \sin \phi$, and the infinitesimal length $db$ becomes $db = 2 a \cos \phi d\phi$. However, once we make this change of variable, we must also enforce $\Vec{c}''_{12}\cdot \apsevector > 0$, so that $\vec{c}_{12}\cdot\hat{a} < 0$, to ensure that the particles move towards each other. On the other hand, we do not need to worry about $b < 2a$ anymore, because the domain of angle $\phi$ automatically imposes that. With this change of variable, total number of collisions becomes $2a f(\Vec{r}_1,\vec{c}_1'',\Vec{\omega}_1'',t)f(\Vec{r}_1,\vec{c}_2'',\Vec{\omega}_2'',t) |\Vec{c}''_{12}| \cos\phi \Theta(\Vec{c}''_{12}\cdot\apsevector) \delta t \delta \vec{c}_1'' \delta \Vec{\omega}_1'' \delta \Vec{c}''_2 \delta \Vec{\omega}''_2 \delta \Vec{r}_1 d\phi $. Furthermore, a more compact expression can be obtained by replacing $|\Vec{c}''_{12}|\cos\phi \Theta(\vec{c}''_{12}\cdot\apsevector)$ with $|\vec{c}''_{12}\cdot\apsevector|\Theta(\vec{c}''_{12}\cdot\apsevector)$.\\\\
Using this expression, the gain term becomes
\begin{align}
    \Gamma^+ \delta\vec{c}_1 \delta \vec{\omega}_1 \delta\vec{r}_1 = 2a\int d\apsevector \int_{R'} d\vec{c}''_1 d\vec{\omega}''_1 \int d\vec{c}''_2 d\vec{\omega}''_2 f(\Vec{r}_1,\Vec{c}''_1, \nonumber\\
    \times \Vec{\omega}''_1,t)f(\Vec{r}_1,\Vec{c}''_2,\Vec{\omega}''_2,t) |\Vec{c}''_{12}| \cos\phi \Theta(\Vec{c}''_{12}\cdot\apsevector) \delta t  \delta \Vec{r}_1,
\end{align}
where $d\apsevector = d\phi$ for a 2D system and $R'$ is the region such that $\vec{c}''_1, \vec{\omega}''_1$ are restricted to yield $\vec{c}_1,\vec{\omega}_1$ after the collision for given $\vec{c}''_2,\vec{\omega}''_2$ and $\apsevector$. 

It is more convenient to express the gain term in terms of target velocities $c_2,\omega_2$. Therefore, we change the integration variables $\vec{c}''_2,\omega_2'' \rightarrow \vec{c}_2,\omega_2$. The Jacobian of this transformation is $1/\beta r$ for our collision model, which gives
\begin{align}
    \Gamma^+ \delta\vec{c}_1 \delta \vec{\omega}_1 \delta\vec{r}_1 = \frac{2a}{\beta r}\int d\apsevector \int_{R} d\vec{c}_1 d\vec{\omega}_1 \int d\vec{c}_2 d\vec{\omega}_2 \nonumber \\ \times f(\Vec{r}_1,\Vec{c}''_1,\Vec{\omega}''_1,t)f(\Vec{r}_1,\Vec{c}''_2,\Vec{\omega}''_2,t) |\Vec{c}''_{12}\cdot\apsevector| \Theta(\Vec{c}''_{12}\cdot\apsevector) \delta t  \delta \Vec{r}_1.
\end{align}
We can furthermore assume that the integration region $R$ is sufficiently small that $\vec{c}_1, \vec{\omega}_1$ integration can be done independently to give $\delta \vec{c}_1 \delta \vec{\omega}_1$ to finally give
\begin{align}
    \Gamma^+ = \frac{2a}{\beta r}\int d\apsevector \int d\vec{c}_2 d\vec{\omega}_2 f(\Vec{r}_1,\Vec{c}''_1,\Vec{\omega}''_1,t)f(\Vec{r}_1,\Vec{c}''_2,\Vec{\omega}''_2,t) \nonumber \\\times 
    |\Vec{c}''_{12}\cdot\apsevector|  \Theta(\Vec{c}''_{12}\cdot\apsevector),
\end{align}
where pre-collision quantities $\vec{c}''_{1,2},\vec{\omega}''_{1,2}$ are functions of $\vec{c}_{1,2}, \vec{\omega}_{1,2}$.

Finally, the loss term can also be rewritten in terms of an integral over $\apsevector$, by considering the infinitesimal collision cylinders and integrating it over the full domain $4a = \int_{-2a}^{2a} db \rightarrow 2a \int d\phi \cos\phi \Theta(-\vec{c}_{12}\cdot\apsevector)  $. Then the full collision term becomes,
\begin{align}
    \mathcal{C}(f_1,f_2) = \Gamma^+ - \Gamma^- = 2a \int d\apsevector d\vec{c}_2 d\vec{\omega_2} |\vec{c}_{12}\cdot\apsevector|\nonumber\\
    \times \Theta(-\vec{c}_{12}\cdot\apsevector) \left[ \frac{f''_1 f''_2}{\beta r^2} - f_1 f_2 \right],     
\end{align}
where we used $\vec{c}''_{12}\cdot\apsevector = -\frac{1}{r}\vec{c}_{12}\cdot\apsevector$, so $\Theta(\vec{c}''_{12}\cdot\hat{a}) = \Theta(-\vec{c}_{12}\cdot\hat{a})$.

\section{Proofs ad Useful Identities} \label{proof_appendix}
In this section, we will prove some statements and obtain certain identities that are useful for the practical calculations that we presented throughout the text, particularly in the following appendices.
\subsection{Collision Operator} \label{appendix:coll_operator}
Let us consider a collision term of the form
\begin{equation}
    \mathcal{C}(f_1,f_2) = \int d\vec{c}_2 d\vec{\omega}_2 d\apsevector|\vec{c}_{12}|\sigma(\vec{c}_{1,2},\vec{\omega}_{1,2},\apsevector) \left[  \frac{f''_1f''_2}{J'} - f_1 f_2   \right],
\end{equation}
where, as always, $''$ denotes the pre-collision quantities. $\sigma(\vec{c}_{1,2},\vec{\omega}_{1,2},\apsevector)$ is the differential cross section of the collision, $\vec{c}_{12} = \vec{c}_1 - \vec{c}_2$ is the shorthand notation for relative linear velocities. We define the coefficient $\frac{1}{J'} \equiv \frac{1}{J K}$, where $J$ is the Jacobian of the transformation between the pre-collision velocities and the current velocities , and $1/K$ is the coefficient that satisfies $\sigma(\vec{c}''_{1,2},\vec{\omega}''_{1,2},\apsevector) = \sigma(\vec{c}_{1,2},\vec{\omega}_{1,2},\apsevector)/K$. $J'$ in general is a function of the integration variables. 

We would like to find a convenient expression for $\mathcal{C}[\psi] \equiv \int d\vec{c}_1 d\omega_1 \mathcal{C}(f_1,f_2) \psi(\vec{c}_1,\omega_1)$, where $\psi$ is any function of center of mass and angular velocities. Due to the symmetry of the exchange of particles, $\mathcal{C}[\psi]$ can be rewritten as
\begin{widetext}
    \begin{equation}
       \mathcal{C}[\psi] = \frac{1}{2} \int d\vec{c}_1 d\vec{\omega}_1 d\vec{c}_2 d\vec{\omega}_2 d\apsevector\sigma(\vec{c}_{1,2},\vec{\omega}_{1,2},\apsevector) \left[  \frac{f''_1f''_2}{J'} - f_1 f_2   \right][\psi(\vec{c}_1,\vec{\omega}_1) + \psi(\vec{c}_2,\vec{\omega}_2)].
\end{equation}
If we then multiply the integration variables and the cross section by $J'$ in the first integral, we obtain, 
\begin{align}
     \mathcal{C}[\psi] = \frac{1}{2}\int d\vec{c}''_1 d\vec{\omega}''_1 d\vec{c}''_2 d\vec{\omega}''_2 d\apsevector \sigma(\vec{c}''_{1,2},\vec{\omega}''_{1,2},\apsevector)f''_1 f''_2 [\psi(\vec{c}_1,\vec{\omega}_1) + \psi(\vec{c}_2,\vec{\omega}_2)] \\
   - \frac{1}{2}\int d\vec{c}_1 d\vec{\omega}_1 d\vec{c}_2 d\vec{\omega}_2 d\apsevector \sigma(\vec{c}_{1,2},\vec{\omega}_{1,2},\apsevector)f_1 f_2 [\psi(\vec{c}_1,\vec{\omega}_1) + \psi(\vec{c}_2,\vec{\omega}_2)].
\end{align}
Now, we will relabel the integration variables in the first integral and simply call pre-collision velocities the current velocities. In that case, the current velocities will become post-collision velocities and we will have,
\begin{align}
    \mathcal{C}[\psi] =  \frac{1}{2}\int d\vec{c}_1 d\vec{\omega}_1 d\vec{c}_2 d\vec{\omega}_2 d\apsevector \sigma(\vec{c}_{1,2},\vec{\omega}_{1,2},\apsevector)f_1 f_2 [\psi(\vec{c}_1',\vec{\omega}_1') + \psi(\vec{c}_2',\vec{\omega}_2') - \psi(\vec{c}_1,\vec{\omega}_1) - \psi(\vec{c}_2,\vec{\omega}_2)] \\
    \equiv  \int d\vec{c}_1 d\vec{\omega}_1 d\vec{c}_2 d\vec{\omega}_2 d\apsevector \sigma(\vec{c}_{1,2},\vec{\omega}_{1,2},\apsevector)f_1 f_2 \Delta \psi,
\end{align}
where $'$ denotes the post-collision quantities.
\end{widetext}
\subsection{Odd-Functions of $\apsevector\times\vec{c}_{12}$} \label{appendix:odd-integrals}
Next we will show that the collision term $\fullcollop[\psi]$ vanishes if $\Delta \psi$ is an odd function of $\apsevector\times\vec{v}_t = \apsevector\times\vec{c}_{12}$, given that the distribution function $f^0$ is of the form as in (\ref{eq:equilibrium_distribution}). Such expressions show up frequently in the calculation of time evolution equations we have in the stability analysis.

\textbf{Proof:} Let us say we have an integral $I$ of the form,
\begin{equation}
    I = \int d\vec{c}_1 d\vec{c}_2 d\omega_1 d\omega_2d\apsevector|\vec{c}_{12}\cdot\apsevector|\Theta(-\vec{c}_{12}\cdot\apsevector)f_1^0 f_2^0 F^o(\apsevector\times\vec{c}_{12}),
\end{equation}
where $F^o(\apsevector\times\vec{c}_{12})$ is any odd function of $\apsevector\times\vec{c}_{12}$. This integral can be rewritten as
\begin{align}
    I = \int d\vec{c}_1 d\vec{c}_2 d\omega_1 d\omega_2 d\apsevector |-\vec{c}_{12}\cdot (\mathcal{R}(\apsevector)[\hat{x}])|  \nonumber \\
    \times \Theta(-\vec{c}_{12}\cdot\mathcal{R}(\apsevector)[\hat{x}]) f_1^0  f_2^0F^o(\apsevector\times\vec{c}_{12}),
\end{align}
where $\mathcal{R}(\apsevector)$ is the rotation matrix that takes $\hat{x}$ to $\apsevector$. Then by using the property of the dot-product, we have
\begin{align}
     I = \int d\vec{c}_1 d\vec{c}_2 d\omega_1 d\omega_2 d\apsevector |-(\mathcal{R}(-\apsevector)[\vec{c}_{12}]\cdot \hat{x})|  \nonumber \\
    \times \Theta(-\mathcal{R}(-\apsevector)[\vec{c}_{12}]\cdot\hat{x}) f_1^0 f_2^0 F^o(\apsevector\times\vec{c}_{12}),
\end{align}
where $\mathcal{R}(-\apsevector)$ is the inverse of $\mathcal{R}(\apsevector)$. If we change the integration variables to $\vec{c}_{1,2} = \mathcal{R}(\apsevector)\vec{c}'_{1,2}$ \footnote{The $'$ used in this subsection has nothing to do with post-collision velocities, which we hope is clear from the context.}, we get
\begin{align}
I = \int d\vec{c}'_1 d\vec{c}'_2 d\omega_1 d\omega_2 d\apsevector |-\vec{c}'_{12}\cdot \hat{x})|  \nonumber \\
    \times \Theta(-\vec{c}'_{12}\cdot\hat{x}) f_1^0 f_2^0 F^o(\apsevector\times\mathcal{R}(\apsevector)[\vec{c}'_{12}]),    
\end{align}
where we used the fact that the Jacobian of the rotation transformation is 1. Finally, using the property of the cross-product in 2D and making another change of variable $\vec{u} = \frac{\vec{c}'_1 - \vec{c}'_2}{\sqrt{2}}$, $\vec{C} = \frac{\vec{c}'_1 + \vec{c}'_2}{\sqrt{2}}$, the integral becomes
\begin{align}
    \label{eq:rotation-trick}
    I = \sqrt{2}\int d\vec{u} d\vec{C} d\omega_1 d\omega_2 d\apsevector |u_x|  \nonumber \\
    \times \Theta(-u_x) f(\vec{u}^2,\vec{C}^2,\omega_1,\omega_2) F^o(\mathcal{R}(-\apsevector)[\apsevector]\times\vec{c}'_{12}) \nonumber\\
    = \sqrt{2}\int d\vec{u} d\vec{C} d\omega_1 d\omega_2 d\apsevector |u_x|  \nonumber \\
    \times \Theta(-u_x) f(\vec{u}^2,\vec{C}^2,\omega_1,\omega_2) F^o(u_y),
\end{align}
where the Jacobian of the given transformation is also 1 and $f(\vec{u}^2,\vec{C}^2,\omega_1,\omega_2) = f_1^0 f_2^0$ in terms of the variables $\vec{u}$ and $\vec{C}$.

Now, inserting $\vec{u}$ and $\vec{C}$ into $f_1^0f_2^0$, and computing $f$ explicitly, we have,
\begin{widetext}
\begin{align}
    f(\vec{u}^2,\vec{C}^2,\omega_1,\omega_2)
    =  n^2 \left(\frac{m}{2\pi T_t}\right)^2 \frac{I}{2\pi T_r} e^{\left[-\frac{m (\vec{u}^2+\vec{C}^2)}{2 T_t} - \frac{I[(\omega_1 - \omega_0)^2 + (\omega_2 - \omega_0)^2 ]}{2 T_r} \right]},
\end{align}
\end{widetext}
and therefore, it is a Gaussian distribution in $\vec{u}$ and $\vec{C}$ as well where $u_x$ and $u_y$ are independent Gaussian variables. Hence, if $F^o$ is an odd function of its argument, the $u_y$ integration becomes 0. This completes the proof.
\begin{widetext}
\subsection{Rotational Invariance of the Collision Operator}
 We would like to show that the collision operator $\mathcal{C}(f,f)$ is rotationally invariant, that is, $\hat{R}\mathcal{C}\hat{R}^{-1}(f,f) = \mathcal{C}(f,f)$, where $\hat{R} \in \mathcal{SO}(2)$ is any rotation matrix on x-y plane. The expression $R\mathcal{C}R^{-1}$, can explicitly be written as,


 \begin{align}
     \hat{R}\mathcal{C}\hat{R}^{-1}(f,f) = \hat{R}\Bigg[\int d\vec{c}_2 d\omega_2 d\apsevector |\vec{c}_{12}\cdot\apsevector| \Theta(-\vec{c}_{12}\cdot\apsevector) \nonumber \\
     \times\left[ \frac{1}{\beta r^2} f( \hat{R}\vec{c}''_1,\omega''_1  ) f( \hat{R}\vec{c}''_2,\omega''_2  ) -  f( \hat{R}\vec{c}_1,\omega_1  ) f( \hat{R}\vec{c}_2,\omega_2  )\right] \Bigg],
 \end{align}
where we used the fact that a scalar function $f$ of any vector $\vec{v}$ transform as $\hat{R}f(\vec{v}) = f(\hat{R}^{-1}\vec{v})$. Now since $\fullcollop[f]$ itself is another function of variables $\vec{c}_1,\omega_1$ only, we furthermore have,

\begin{align}
\hat{R}\mathcal{C}\hat{R}^{-1}(f,f) = \Bigg[\int d\vec{c}_2 d\omega_2 d\apsevector |(\hat{R}^{-1}\vec{c}_1 - \vec{c}_2 )\cdot\apsevector| \Theta(-(\hat{R}^{-1}\vec{c}_1 - \vec{c}_2)\cdot\apsevector) \nonumber \\
     \times\left[ \frac{1}{\beta r^2} f( \hat{R}\vec{c}''_1(\hat{R}^{-1}\vec{c}_1...),\omega''_1(\hat{R}^{-1}\vec{c}_1...)  ) f( \hat{R}\vec{c}''_2(\hat{R}^{-1}\vec{c}_1...),\omega''_2(\hat{R}^{-1}\vec{c}_1...)  ) -  f( \vec{c}_1,\omega_1  ) f( \hat{R}\vec{c}_2,\omega_2  )\right] \Bigg],    
\end{align}
where inside $f$, ellipses denote that the function $f$ is in general a function of all variables $\vec{c}_2,\omega_1,\omega_2,\apsevector$ but $\hat{R}$ only acts on $\vec{c}_1$. Next, we do the change in integration variable $\vec{c}_2 \rightarrow \hat{R}^{-1} \vec{c}_2, \apsevector \rightarrow \hat{R}^{-1}\apsevector$, and since the Jacobians of these transformations are 1, we get,
\begin{align}
\hat{R}\mathcal{C}\hat{R}^{-1}(f,f) = \Bigg[\int d\vec{c}_2 d\omega_2 d\apsevector |(\hat{R}^{-1}\vec{c}_1 - \hat{R}^{-1}\vec{c}_2 )\cdot \hat{R}^{-1} \apsevector| \Theta(-(\hat{R}^{-1}\vec{c}_1 - \hat{R}^{-1}\vec{c}_2)\cdot\hat{R}^{-1}\apsevector) \nonumber \\
     \times\bigg[ \frac{1}{\beta r^2} f( \hat{R}\vec{c}''_1(\hat{R}^{-1}(\vec{c}_1,\vec{c}_2,\apsevector)...),\omega''_1(\hat{R}^{-1}(\vec{c}_1,\vec{c}_2,\apsevector)...)  ) f( \hat{R}\vec{c}''_2(\hat{R}^{-1}(\vec{c}_1,\vec{c}_2,\apsevector)...),\omega''_2(\hat{R}^{-1}(\vec{c}_1,\vec{c}_2,\apsevector)...)  ) \nonumber \\
     -  f( \vec{c}_1,\omega_1  ) f( \vec{c}_2,\omega_2  )\bigg] \Bigg],    
\end{align}
where $\hat{R}^{-1}(\vec{c}_1,\vec{c}_2,\apsevector)$ is a shorthand notation for $(\hat{R}^{-1}\vec{c}_1,\hat{R}^{-1}\vec{c}_2,\hat{R}^{-1}\apsevector)$.
Finally, we note from equations \ref{eq:post-vel1}-\ref{eq:post-rot-vel} that $\vec{c}'_{1,2}(\hat{R}\vec{c}_1,\hat{R}\vec{c}_2,\hat{R}\apsevector,\omega_1,\omega_2) = \hat{R}\vec{c}_{1,2}'(\vec{c}_1,\vec{c}_2,\apsevector,\omega_1,\omega_2) $ and $\omega'_{1,2}(\hat{R}\vec{c}_1,\hat{R}\vec{c}_2,\hat{R}\apsevector,\omega_1,\omega_2) = \omega_{1,2}'(\vec{c}_1,\vec{c}_2,\apsevector,\omega_1,\omega_2)$, which also holds for pre-collision velocities. Because of that and the fact that rotations do not change dot-products, we have,
\begin{align}
\hat{R}\mathcal{C}\hat{R}^{-1}(f,f) = \Bigg[\int d\vec{c}_2 d\omega_2 d\apsevector |\vec{v}_{12}\cdot\apsevector| \Theta(-\vec{v}_{12}\cdot\apsevector) \nonumber \\
     \times\left[ \frac{1}{\beta r^2} f( \vec{c}''_1,'\omega_1  ) f( \vec{c}''_2,'\omega_2  ) -  f( \vec{c}_1,\omega_1  ) f( \vec{c}_2,\omega_2  )\right] \Bigg] = \mathcal{C}(f_1,f_2),   
\end{align}
and therefore collision operator is rotationally invariant. Repeating the same argument, we can also show that the linear collision operator $\linearcollop$ is also rotationally invariant.
\end{widetext}

\section{Stability Analysis}
In this appendix, we will derive the time evolution equations for the variables $\omega_0$, $T_t$ and $T_r$, and show that the assumption of a steady state Gaussian distribution is self-consistent. We will do that in three steps: We will first find the fixed point of the time evolution equations. Then we will obtain the dimensionless versions of them for computational convenience. Finally, we will perform a linear stability analysis around this fixed point and show that it is stable.
\subsection{Derivation of Time Evolution Equations}
We will start with the Boltzmann equation \eqref{eq:Boltzmann_eq} assuming the spatially homogeneous Gaussian distribution of the form in Eq. \eqref{eq:equilibrium_distribution}, which is,
\begin{equation}
    \partial_t f + d\partial_\omega f = \mathcal{C}(f,f).
\end{equation}
To obtain the evolution equation for the average angular velocity, we multiply this equation by $\omega/n$ and integrate it over the full velocity phase-space, which gives,
\begin{equation}
    \partial_t \omega_0 - d = \frac{1}{n}\mathcal{C}[\omega],
\end{equation}
where for the first term on the left-hand side we simply pulled-out the time derivative out of the integral and used the definition of averages, and for the second term, we performed integration by parts. As we showed in the App. \ref{appendix:coll_operator}, the right-hand side becomes
\begin{align}
    \mathcal{C}[\omega] = \frac{2a}{n} \int d\vec{c}_1 d\vec{c}_2 d\omega_1 d\omega_2 d\apsevector |\vec{c}_{12}\cdot\apsevector| \nonumber \\
    \times \Theta(-\vec{c}_{12}\cdot\apsevector)f_1^0f_2^0\Delta \omega,
\end{align}
where, calculating from the collision rules in Eq. \eqref{eq:post-vel1} - \eqref{eq:post-rot-vel}, $\Delta \omega$ becomes,
\begin{equation}
    \Delta \omega = \frac{\modtangentfric}{aq} \apsevector\times\vec{v}_t - \frac{\modtangentfric}{q} (\omega_1 + \omega_2).
\end{equation}
We proved in the App. \ref{appendix:odd-integrals} that the odd functions of $\vec{c}_{12}\times\apsevector = \vec{v}_t \times \apsevector$ are integrated to 0 and therefore the first term does not contribute. The collision term $\fullcollop[\omega]$ then becomes
\begin{align}
     \mathcal{C}[\omega] = -\frac{2a\modtangentfric}{qn}\int d\vec{c}_1 d\vec{c}_2 d\omega_1 d\omega_2 |-\vec{c}_{12}\cdot\apsevector| \nonumber \\
     \times \Theta(-\vec{c}_{12}\cdot\apsevector) f_1 f_2 (\omega_1 + \omega_2).
\end{align}
Using the rotation argument in Eq. \eqref{eq:rotation-trick}, this expression becomes
\begin{align}
    \mathcal{C}[\omega] = -\frac{2\sqrt{2} a \modtangentfric}{qn} \int d\vec{u} d\vec{C} d\omega_1 d\omega_2 d\apsevector|u_x|\nonumber \\
    \times \Theta(-u_x)f(\vec{u}^2,\vec{C}^2,\omega_1,\omega_2)(\omega_1 + \omega_2),
\end{align}
which is simply a multiplication of independent Gaussian averages after the trivial $\apsevector$ integral is performed, i.e.,
\begin{align}
    \mathcal{C}[\omega] = -\frac{4\sqrt{2} a n \modtangentfric \pi }{q} \langle |u_x|\Theta(-u_x) \rangle \langle (\omega_1 + \omega_2) \rangle \nonumber\\
    = -G \frac{\modtangentfric}{q} \omega_0 T_t^{1/2},
 \end{align}
 where $G = \frac{8 a n \sqrt{\pi}}{\sqrt{m}}$ and $\langle \rangle$ denotes average over Gaussian distribution. Combining all of them, the equation for rotational velocity becomes,
 \begin{equation}
     \partial_t \omega_0 = d - G \frac{\modtangentfric}{q} \omega_0 T_t^{1/2}.
 \end{equation}
 We can repeat the same procedure for translational $T_t$ and rotational $T_r$ temperatures by multiplying the Boltzmann equation by $\frac{m \vec{c}_1^2}{2}$ and $I(\omega_1 - \omega_0)$ respectively. For the translational energy, the collision term becomes,
 \begin{align}
     \mathcal{C}\left[ \frac{m \vec{c}^2}{2} \right] = ma\int d\vec{c}_1 d\vec{c}_2 d\omega_1 d\omega_2 d\apsevector |\vec{c}_{12}\cdot\apsevector| \nonumber \\
     \times \Theta(-\vec{c}_{12}\cdot\apsevector) f_1^0 f_2^0 \Delta \vec{c}^2,
 \end{align}
 where,
 \begin{equation}
     \Delta \vec{c}^2 = -\frac{1-r^2}{4}\vec{v}_n^2 - \modtangentfric(1-\modtangentfric)\vec{v}_t^2 + \modtangentfric^2 \vec{v}_r^2 + F^o(\vec{v}_t),
 \end{equation}
 and $F^o(\vec{v}_t)$ is an odd-function of $\vec{v}_t$ and hence of $\apsevector\times\vec{c}_{12}$ and does not contribute to the collision term. Using the rotational argument and computing the Gaussian averages, the full time evolution equation becomes,
 \begin{equation}
     \partial_t T_t = -G A_1 T_t^{3/2} + G A_2 T_t^{1/2} T_r + GA_2 T_t^{1/2} (2I\omega_0^2),
 \end{equation}
where $A_1 = \left[ \frac{1-r^2}{4} + \frac{\modtangentfric(1-\modtangentfric)}{2} \right]$, $A_2 = \frac{\modtangentfric^2}{2q}$.
 Finally, the collision term for the rotational temperature is,
 \begin{align}
     \mathcal{C}[I(\omega - \omega_0)^2] = 2a \int d\vec{c}_1 d\vec{c}_2 d\omega_1 d\omega_2 d\apsevector |\vec{c}_{12}\cdot\apsevector| \nonumber \\ \times \Theta(-\vec{c}_{12}\cdot\apsevector) f_1^0 f_2^0 I \Delta (\omega -\omega_0)^2,
 \end{align}
 where,
 \begin{align}
      \Delta (\omega - \omega_0)^2 = \frac{\modtangentfric^2}{a^2 q^2} (\apsevector\times\vec{v}_t)^2 + \frac{\modtangentfric^2}{q^2}(\omega_1 + \omega_2)^2 \\
      - \frac{\modtangentfric}{q}(\omega_1 + \omega_2 - 2\omega_0)(\omega_1 + \omega_2) + F^o(\apsevector\times\vec{v}_t).
 \end{align}
 Carrying out the full integral yields,
 \begin{equation}
    \partial_t{T_r} = 2GA_2 T_t^{3/2} - 2 G B_1 T_r T_t^{1/2} + 2GB_2 T_t^{1/2}(2I{\omega}_0^2),
\end{equation}
where $    A_2 =  \frac{\modtangentfric^2}{2q}, B_1 = \frac{\modtangentfric}{2q}\left( 1 - \frac{\modtangentfric}{q} \right)$ and $B_2 = \frac{\modtangentfric^2}{2q^2}$. These are the equations given in the main text.
\subsection{Non-Dimensionalization}
\label{app:dimless_quant}
Now, we will introduce a natural temperature and time scales of the system to obtain the dimensionless angular velocity $\Tilde{\omega}_0$ and temperatures $\Tilde{T}_t$ and $\Tilde{T}_r$. Because the only time scale in the system is the external torque $d$, we do not have too many options. We will simply choose the dimensionless angular velocity as $\Tilde{\omega}_0 \equiv \omega_0/\sqrt{|d|}$ and the temperatures as $\Tilde{T}_{t,r} \equiv \frac{T_{t,r}}{ma^2|d|} $. Then the only dimensionful quantity left in the equations is the factor $G = \frac{8\sqrt{\pi} n a}{\sqrt{m}}$. For a 2D system, we can simply define $\Tilde{G} \equiv \sqrt{m} a G$. Using these definitions and the equilibrium solutions \eqref{eq:omega_equilibrium} - \eqref{eq:tr_T_equilibrium} we have in the main text, we have the dimensionless equilibrium solutions as,
\begin{align}
    \Tilde{\omega}_0 &= \frac{2\text{sgn}(d)}{C_1 \Tilde{G} \Tilde{T_t}^{1/2}}\\
    \Tilde{T}_r &= \frac{A_1}{A_2} \Tilde{T}_t - \frac{8 q}{\Tilde{G}^2 C_1^2 \Tilde{T}_t}\\
    \Tilde{T}_t &= \left[  \frac{  \frac{8q}{C_1^2 \Tilde{G}^2}(B_1 + B_2)   } { \frac{B_2 A_1}{A_2} - A_2 }  \right]^{1/2}.
\end{align}
The signs of these temperatures depend only on the friction parameters $\restcoefnormal$ and $\modtangentfric$, and numerically plotting them over the domain of allowed values $\restcoefnormal \in [0,1], \modtangentfric \in [0,1/3]$ shows that they are always positive.
\subsection{Stability Analysis}
\label{app: Stability Analysis}
To check if these values are indeed equilibrium solutions, we also need to show that they are a stable fixed point of the Eqs. \eqref{eq:omega_equilibrium}-\eqref{eq:tr_T_equilibrium}. If we non-dimensionalize the evolution equations with the same procedure, we get
\begin{widetext}
\begin{align*}
    \partial_{\Tilde{t}} \Tilde{\omega}_0 &= \text{sgn}(d) - \frac{\Tilde{G}C_1}{2}\Tilde{\omega}_0 \Tilde{T}_t^{1/2} \equiv f(\Tilde{\omega}_0,\Tilde{T}_t,\Tilde{T}_r) \\
    \partial_{\Tilde{t}} \Tilde{T}_t &= - \Tilde{G} A_1 \Tilde{T}^{3/2}_t + \Tilde{G} A_2 \Tilde{T}_t^{1/2} \Tilde{T}_r + 2\Tilde{G} A_2 \Tilde{T}_t^{1/2}  \Tilde{\omega}_0^2 q \equiv g(\Tilde{\omega}_0,\Tilde{T}_t,\Tilde{T}_r) \\
    \partial_{\Tilde{t}} \Tilde{T}_r &= 2\Tilde{G} A_2 \Tilde{T}^{3/2} - 2\Tilde{G} B_1 \Tilde{T}_r \Tilde{T}_t^{1/2} + 4\Tilde{G} B_2 \Tilde{T}_t^{1/2}  q \Tilde{\omega}_0^2 \equiv h(\Tilde{\omega}_0,\Tilde{T}_t,\Tilde{T}_r).
\end{align*}
\end{widetext}
The fixed point will be stable if the eigenvalues of the matrix,

\begin{equation}
    \begin{pmatrix}
        \frac{\partial f}{\partial \Tilde{\omega}_0} & \frac{\partial f}{\partial \Tilde{T}_t} & \frac{\partial f}{\partial \Tilde{T}_r } \\
         \frac{\partial g}{\partial \Tilde{\omega}_0} & \frac{\partial g}{\partial \Tilde{T}_t} & \frac{\partial g}{\partial \Tilde{T}_r } \\
          \frac{\partial h}{\partial \Tilde{\omega}_0} & \frac{\partial h}{\partial \Tilde{T}_t} & \frac{\partial h}{\partial \Tilde{T}_r }
    \end{pmatrix},
\end{equation}
evaluated at the fixed point has negative eigenvalues. Sweeping all the parameters $\restcoefnormal \in [0,1]$ and $\modtangentfric \in [0,1/3]$ and testing it on several $\Tilde{G}$ values numerically suggests that the solution is indeed stable.

\section{Molecular Dynamics Simulation}
\label{app:MD-sim}
Using the Julia Agents package, we simulate a system of 400 disks with radius $a = 0.02$ on a 4 x 4 square with periodic boundary conditions. We start the simulation by assigning random angular and linear momenta to each particle, chosen from a normal distribution that is different from the expected equilibrium distribution. We then run the simulation with the collision rules described in equations \ref{eq:post-vel1}-\ref{eq:post-rot-vel}. After the system reaches a steady state, we sample linear and angular velocities over a total of 20000 time steps. 

\begin{figure*}[b]
\includegraphics[scale=0.45]{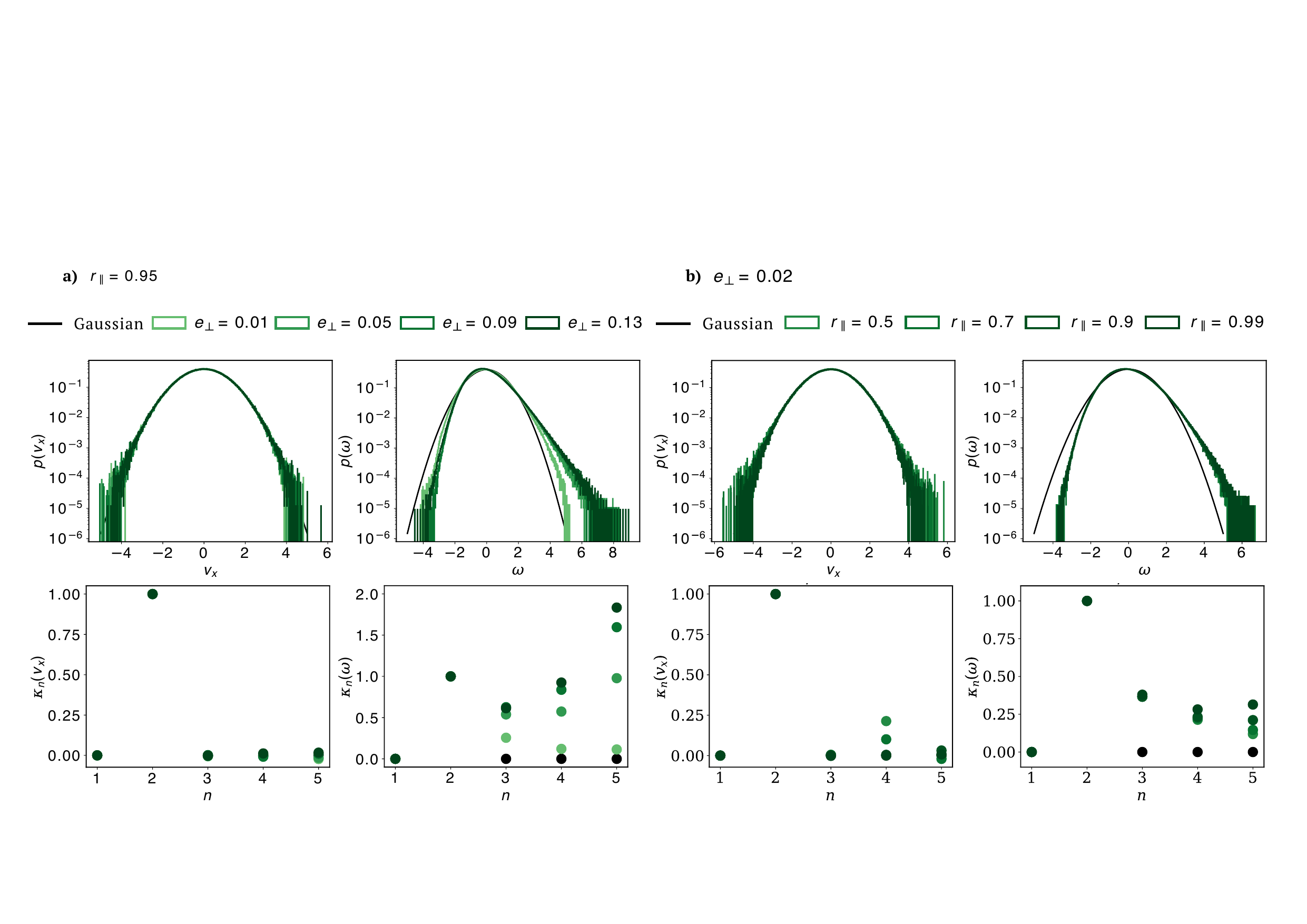}
\caption{\textbf{Distributions of Velocities:} Distributions of center of mass $v_x$ and angular $\omega$ velocities and their cumulants $\kappa_n$ obtained from molecular dynamics simulations. The distributions of angular velocities are normalized by subtracting average angular velocity $\omega_0$ from all terms and dividing each term by the standard deviation $\sigma_\omega$. Velocity distributions are normalized by dividing each term by the standard deviation $\sigma_{v_x}$. A Gaussian distribution with 0 mean and standard deviation 1 is also plotted for reference. \textbf{a)} For a fixed value of normal restitution $\restcoefnormal = 0.95$, distributions of the x component of linear velocity $v_x$ and angular velocity $\omega$ is given for several tangential restitution coefficients $\modtangentfric$. Furthermore, cumulants $\kappa_n$ corresponding to each case are plotted. The distributions of center of mass velocity are not significantly affected by the change in tangential restitution and remain fairly close to a Gaussian distribution. On the other hand, distributions of angular velocities depend heavily on the tangential restitution and significantly deviate from a Gaussian as the tangential restitution is increased, with a visible skewness. For sufficiently small tangential restitution, Gaussian can still be a good approximation. \textbf{b)} Same analysis in \textbf{a)} is repeated by fixing the tangential restitution $\modtangentfric = 0.02$ and varying the normal restitution coefficient $\restcoefnormal$. Distributions of angular velocities are not significantly affected by the change in normal restitution. The distributions of center of mass velocities are affected, the distribution still remains fairly close to a Gaussian even for small $\restcoefnormal$ values. }
\label{fig:velocity_distributions_MD}
\end{figure*}

The distributions of linear and angular velocities obtained from the simulations are given in Fig. \ref{fig:velocity_distributions_MD}, in log scale on the y axis. Each distribution is divided by its standard deviation and shifted by its mean, and plotted together with a reference Gaussian of standard deviation 1 and 0 mean. We further plot the absolute values of the cumulants of the distributions up to the sixth cumulant.

We observe that the distribution of linear velocities remains close to a Gaussian distribution, while the rotational velocities can significantly deviate from one, with a visible tilt. Furthermore, the distribution of angular velocities is affected only slightly by the increase in the normal restitution coefficient $\restcoefnormal$, while a change in the tangential restitution coefficient $\modtangentfric$ can affect it significantly. 

The deviation from the Gaussian distribution of linear velocities in a driven granular gas is observed in studies with different driving methods \cite{puglisi2002fluctuation, cafiero2000two}. We observe a qualitatively similar behavior. The skewness of the angular velocity distributions is expected, as the torque in a given direction favors one side of the curve with respect to the mean.


\section{Full Hydro via Adiabatic Elimination}
\label{app:full_hydro_adiabatic}
In this section, we will explicitly write down the equation $\partial_t (\Phi_\alpha,\phi) = [L_{eff}]_{\alpha \beta}(\Phi_\beta,\phi)$, for all $\Phi_\alpha \in \mathcal{S}$. Although some terms can be discarded with symmetry arguments without any calculation, most of the terms are very difficult to compute analytically, and we will leave them as inner products to be computed numerically when needed.

Some terms can be proved to be 0 using the rotational invariance of the collision operator and the fact that some variables are collision invariants. Apart from these, some inner products generally turn out to be zero, and how we decide on them is as follows: We numerically calculate those inner products across different $\modtangentfric$ values. If the results do not have a clear pattern and fluctuate around 0, then we take them to be 0. 
\begin{widetext}
    After discarding the 0 terms as described above and using the definitions \eqref{eq:perturbation_as_inner_prod}, the full hydrodynamic equations can be obtained as, for the mass density,
    \begin{subequations}
        \begin{equation}
        \label{eq:adibatic_hydro_1}
            \textcolor{convection_color}{D_t \delta \rho + \rho^{(0)} \nabla \cdot \delta \vec{u}} = 0
        \end{equation}
    \begin{align}
        \label{eq:Boltzmann_hydro_mass}
        \textcolor{convection_color}{\partial_t \delta \rho  + i q_i \rho^{(0)} \delta v_i} = 0,
           \end{align}
    \end{subequations}
           for the velocity field,
    \begin{subequations}   
    \begin{equation}
        \textcolor{convection_color}{\rho^{(0)} D_t \delta u_i }\mathbin{\textcolor{relaxation_color}{+}}{\textcolor{relaxation_color}{  \partial_jP_{ij}^{(rel)}}}    \mathbin{\textcolor{transport_color}{+}}  \textcolor{transport_color}{ \partial_j P_{ij}^{(visc)}} = 0
    \end{equation}
           \begin{align}
        \textcolor{convection_color}{\sqrt{\beta_t m}\partial_t \delta u_i =} 
    \mathbin{\textcolor{relaxation_color}{-}}   \textcolor{relaxation_color}{iq_j\delta_{i,j}\left[ \frac{\delta \rho}{\rho\sqrt{\beta_t m}} + \frac{\delta T_t}{T_t \sqrt{\beta_t m}} \right]  }  \nonumber\\ \mathbin{\textcolor{transport_color}{-}}
        \textcolor{transport_color}{q_j q_k ( v^j_{FS} \Phi_{v,i}, L_{FF}^{-1} v^k_{FS}\Phi_{v,\ell} ) \sqrt{\beta_t m}\delta u_\ell}, 
    \end{align}
    \end{subequations}
    for the angular velocity field,
    \begin{subequations}
        \begin{equation}
            \textcolor{convection_color}{I n^{(0)} D_t \delta \omega} \mathbin{\textcolor{relaxation_color}{+}}\textcolor{relaxation_color}{ \partial_i M_i^{(rel)}} \mathbin{\textcolor{transport_color}{+}} \textcolor{transport_color}{\partial_i M^{TP}_i} \mathbin{\textcolor{external_force_color}{-}} \textcolor{external_force_color}{I d \delta n} = \textcolor{collision_color}{\mathcal{C}^{(1)}[I \omega]},
        \end{equation}
    
    \begin{align}
    \textcolor{convection_color}{\sqrt{\beta_r I}\partial_t \delta \omega} \mathrel{\textcolor{convection_color}{=} }  \textcolor{external_force_color}{(L^\dagger[\Phi_\omega], \Phi_\rho) \frac{\delta \rho}{\rho} } \mathbin{\textcolor{collision_color}{+} }  \textcolor{collision_color}{\sqrt{\beta_r I}(L^\dagger[\Phi_\omega], \Phi_\omega)\delta \omega + (L^\dagger[\Phi_\omega], \Phi_{et})\frac{\delta T_t}{T_t} }\nonumber \\
     \textcolor{transport_color}{-q_i q_j \left[ \sqrt{\beta_r I}(\Phi_\omega,v^i_{SF}L^{'-1}_{FF} v^j_{FS}\Phi_\omega)\delta\omega 
 +  (\Phi_\omega,v^i_{SF}L^{'-1}_{FF} v^j_{FS}\Phi_{et})\frac{\delta T_t}{T_t} + (\Phi_\omega,v^i_{SF}L^{-1}_{FF} v^j_{FS}\Phi_{er})\frac{\delta T_r}{\sqrt{2}T_r} \right] } \nonumber \\
    \textcolor{relaxation_color}{-iq_i \sqrt{\beta_t m }( Q v^\mu \Phi_\omega, L^{-1}_{FF} L_{FS} \Phi_{v,j} )\delta v_j },
\end{align}
\end{subequations}
for the translational temperature,
\begin{subequations}
    
\begin{equation}
    \textcolor{convection_color}{n^{(0)} D_t \delta e_t} \mathbin{\textcolor{transport_color}{+} } \textcolor{transport_color}{\nabla \cdot \Vec{Q}^{(TP)}} \mathbin{\textcolor{relaxation_color}{+} } \textcolor{relaxation_color}{\nabla \cdot Q^{(rel)}} \mathrel{\textcolor{collision_color}{=} } \textcolor{collision_color}{\mathcal{C}^{(1)}[mv^2/2]}
\end{equation}
\begin{align}
\textcolor{convection_color}{\frac{1}{T_t}\partial_t \delta T_t =}  \textcolor{collision_color}{\sqrt{\beta_r I}(L^\dagger[\Phi_{et}], \Phi_\omega)\delta \omega + (L^\dagger[\Phi_{et}], \Phi_{et})\frac{\delta T_t}{T_t} + (L^\dagger[\Phi_{et}], \Phi_{er}) \frac{\delta T_r}{\sqrt{2} T_r}}\nonumber \\
     \textcolor{transport_color}{-q_i q_j\left[ \sqrt{\beta_r I}(\Phi_{et},v^i_{SF}L^{-1}_{FF} v^j_{FS}\Phi_\omega)\delta\omega 
+  (\Phi_{et},v^i_{SF}L^{-1}_{FF} v^j_{FS}\Phi_{et})\frac{\delta T_t}{T_t} + (\Phi_{et},v^i_{SF}L^{-1}_{FF} v^j_{FS}\Phi_{er})\frac{\delta T_r}{\sqrt{2}T_r} \right] }\nonumber \\
\textcolor{convection_color}{-iq_i \frac{\delta v_i}{\sqrt{\beta_t m}} } \mathbin{\textcolor{relaxation_color}{-}} \textcolor{relaxation_color}{ iq_i \sqrt{\beta_t m }( Q v^i \Phi_{et}, L^{-1}_{FF} L_{FS} \Phi_{v,\nu} )\delta v_\nu},
\end{align}
\end{subequations}
and for the rotational temperature,
\begin{subequations}
\begin{equation}
     \textcolor{convection_color}{n D_t \delta e_r} \mathbin{\textcolor{transport_color}{+} } \textcolor{transport_color}{\nabla \cdot \Vec{Q}_\omega^{(TP)}} \mathbin{\textcolor{relaxation_color}{+} } \textcolor{relaxation_color}{\nabla \cdot \Vec{Q}_\omega^{(rel)}} \mathrel{\textcolor{collision_color}{=} } \textcolor{collision_color}{\mathcal{C}^{(1)}[I\omega^2/2 - I\omega \Bar{\omega}]}
\end{equation}
\begin{align}
\label{eq:adibatic_hydro_fin}
\textcolor{convection_color}{\frac{1}{\sqrt{2}T_r}\partial_t \delta T_r = } \textcolor{collision_color}{\sqrt{\beta_r I}(L^\dagger[\Phi_{er}], \Phi_\omega)\delta \omega + (L^\dagger[\Phi_{er}], \Phi_{er})\frac{\delta T_t}{T_t} + (L^\dagger[\Phi_{et}], \Phi_{er}) \frac{\delta T_r}{\sqrt{2}T_r} } \nonumber \\
   \textcolor{transport_color}{ -q_i q_j \left[ \sqrt{\beta_r I}(\Phi_{er},v^i_{SF}L^{-1}_{FF} v^j_{FS}\Phi_\omega)\delta\omega 
 +  (\Phi_{er},v^i_{SF}L^{-1}_{FF} v^j_{FS}\Phi_{et})\frac{\delta T_t}{T_t} + (\Phi_{er},v^i_{SF}L^{-1}_{FF} v^j_{FS}\Phi_{er})\frac{\delta T_r}{\sqrt{2}T_r} \right] } \nonumber \\
\textcolor{relaxation_color}{-iq_i \sqrt{\beta_t m }( Q v^i \Phi_{er}, L^{-1}_{FF} L_{FS} \Phi_{v,j} )\delta v_j}.
    \end{align}
    \end{subequations}
\end{widetext}
As can be seen, any hydrodynamic density $\rho_A$ obeys the equation of the form
\begin{equation}
    \textcolor{convection_color}{\partial_t \delta \rho_A}  \mathbin{\textcolor{relaxation_color}{+}} \textcolor{relaxation_color}{\nabla \cdot J^{(rel)}} \mathbin{\textcolor{transport_color}{+}} \textcolor{transport_color}{\nabla \cdot J^{(TP)}} \mathbin{\textcolor{external_force_color}{+}} \textcolor{external_force_color}{f^{(ext)}} \mathbin{\textcolor{collision_color}{+}} \textcolor{collision_color}{S^{(sink)}} = 0,
\end{equation}
where \textcolor{convection_color}{$\partial_t \delta \rho_A + iq_i \rho_A^{(0)}\delta u_i$ } denotes the terms due to \textcolor{convection_color}{continuity}, \textcolor{relaxation_color}{$ J^{(rel)  }\sim \delta \rho_i $ } denotes \textcolor{relaxation_color}{relaxation currents } proportional to the local deviation from the steady state, \textcolor{transport_color}{$J^{(TP)} \sim \nabla \delta \rho_i $} denotes the \textcolor{transport_color}{transport currents} proportional to the gradients of hydrodynamic densities, \textcolor{external_force_color}{$f^{(ext)}$} denotes the \textcolor{external_force_color}{external forces}, and finally \textcolor{collision_color}{$S^{(sink)}$} denotes the \textcolor{collision_color}{loss terms due to collisions}.

These terms are color-coded in Eqs. \eqref{eq:adibatic_hydro_1}-\eqref{eq:adibatic_hydro_fin}. As can be seen, the mass conservation equation does not change, while all other equations have both relaxational and transport current terms that respect the symmetries of the system. As expected, equations that correspond to non-conserved quantities ($\delta \omega, \delta T_t, \delta T_r$) have collisional loss terms and the external driving term also appears in the angular momentum equation. 
\begin{widetext}

\subsection{Computation of Current Terms}

\label{app:other_transport_coeffs}
As mentioned in the main text, we list the computational expressions for all the non-zero relaxation and transport coefficients in terms of the non-dimensional inner products. The details on how they are computed are given in App. \ref{app:Computational_Details}.

Expressions for non-zero relaxation currents read,
\begin{align}
    \label{eq:all_relaxation_coef}
    \tilde{\zeta}^o_{\omega,v} &= \frac{1}{2}\left[ (Q\tilde{v}_x\Phi_\omega, \tilde{L}^{-1}_{FF} L_{FS} \Phi_{v,x} )_{ND} + (Q\tilde{v}_y\Phi_\omega, \tilde{L}^{-1}_{FF} L_{FS} \Phi_{v,y} )_{ND}\right] \nonumber \\
    \tilde{\zeta}_{\omega,v} & = \frac{1}{2}\left[ (Q\tilde{v}_x\Phi_\omega, \tilde{L}^{-1}_{FF} L_{FS} \Phi_{v,y} )_{ND} - (Q\tilde{v}_y\Phi_\omega, \tilde{L}^{-1}_{FF} L_{FS} \Phi_{v,x} )_{ND}\right] \nonumber \\
    \tilde{\zeta}_{T_t,v} &= \frac{1}{2}\left[ (Q\tilde{v}_x\Phi_{et}, \tilde{L}^{-1}_{FF} L_{FS} \Phi_{v,x} )_{ND} + (Q\tilde{v}_y\Phi_{et}, \tilde{L}^{-1}_{FF} L_{FS} \Phi_{v,y} )_{ND}\right] \nonumber \\
    \tilde{\zeta}^o_{T_t,v} & = \frac{1}{2}\left[ (Q\tilde{v}_x\Phi_{et}, \tilde{L}^{-1}_{FF} L_{FS} \Phi_{v,y} )_{ND} - (Q\tilde{v}_y\Phi_{et}, \tilde{L}^{-1}_{FF} L_{FS} \Phi_{v,x} )_{ND}\right] \nonumber \\
    \tilde{\zeta}_{T_r,v} &= \frac{1}{2}\left[ (Q\tilde{v}_x\Phi_{er}, \tilde{L}^{-1}_{FF} L_{FS} \Phi_{v,x} )_{ND} + (Q\tilde{v}_y\Phi_{er}, \tilde{L}^{-1}_{FF} L_{FS} \Phi_{v,y} )_{ND}\right] \nonumber \\
    \tilde{\zeta}^o_{T_r,v} & = \frac{1}{2}\left[ (Q\tilde{v}_x\Phi_{er}, \tilde{L}^{-1}_{FF} L_{FS} \Phi_{v,y} )_{ND} - (Q\tilde{v}_y\Phi_{er}, \tilde{L}^{-1}_{FF} L_{FS} \Phi_{v,x} )_{ND}\right],
\end{align}
and for transport coefficients, we have,
\begin{align}
    \label{eq:all_transport_coef}
    \tilde{\eta} &=  - \frac{1}{4} \Big[ (Q\tilde{v}_x\Phi_{v,x}, \tilde{L}^{'-1}_{FF} Q \tilde{v}_x \Phi_{v,x} )_{ND} + (Q\tilde{v}_y\Phi_{v,y}, \tilde{L}^{'-1}_{FF} Q \tilde{v}_y \Phi_{v,y} )_{ND} \nonumber \\ &- (Q\tilde{v}_y\Phi_{v,y}, \tilde{L}^{'-1}_{FF} Q \tilde{v}_x \Phi_{v,x} )_{ND} - (Q\tilde{v}_x\Phi_{v,x}, \tilde{L}^{'-1}_{FF} Q \tilde{v}_y \Phi_{v,y} )_{ND}   \Big] \nonumber \\
    \tilde{\eta}^o &=  - \frac{1}{4} \Big[ (Q\tilde{v}_x\Phi_{v,x}, \tilde{L}^{'-1}_{FF} Q \tilde{v}_x \Phi_{v,y} )_{ND} + (Q\tilde{v}_x\Phi_{v,x}, \tilde{L}^{'-1}_{FF} Q \tilde{v}_y \Phi_{v,x} )_{ND} \nonumber \\ &- (Q\tilde{v}_y\Phi_{v,y}, \tilde{L}^{'-1}_{FF} Q \tilde{v}_x \Phi_{v,y} )_{ND} - (Q\tilde{v}_y\Phi_{v,y}, \tilde{L}^{'-1}_{FF} Q \tilde{v}_y \Phi_{v,x} )_{ND}   \Big] \nonumber \\
    \tilde{\eta}_\omega &= - \frac{1}{2} \left[ (Q\tilde{v}_x\Phi_\omega, \tilde{L}^{'-1}_{FF} Q \tilde{v}_x \Phi_\omega )_{ND} + (Q\tilde{v}_y\Phi_\omega, \tilde{L}^{'-1}_{FF} Q \tilde{v}_y \Phi_\omega )_{ND}  \right] \nonumber \\
    \tilde{\eta}^o_\omega &= - \frac{1}{2} \left[ (Q\tilde{v}_x\Phi_\omega, \tilde{L}^{'-1}_{FF} Q \tilde{v}_y \Phi_\omega )_{ND} - (Q\tilde{v}_y\Phi_\omega, \tilde{L}^{'-1}_{FF} Q \tilde{v}_x \Phi_\omega )_{ND}  \right] \nonumber \\
    \tilde{\Omega}_{\omega, T_t} &= - \frac{1}{2} \left[ (Q\tilde{v}_x\Phi_\omega, \tilde{L}^{'-1}_{FF} Q \tilde{v}_y \Phi_{et} )_{ND} - (Q\tilde{v}_y\Phi_\omega, \tilde{L}^{'-1}_{FF} Q \tilde{v}_x \Phi_{et} )_{ND}  \right] \nonumber \\
\tilde{\Omega}^o_{\omega, T_t} &= - \frac{1}{2} \left[ (Q\tilde{v}_x\Phi_\omega, \tilde{L}^{'-1}_{FF} Q \tilde{v}_x \Phi_{et} )_{ND} + (Q\tilde{v}_x\Phi_\omega, \tilde{L}^{'-1}_{FF} Q \tilde{v}_x \Phi_{et} )_{ND}  \right] \nonumber \\
\tilde{\Omega}_{\omega, T_r} &= - \frac{1}{2} \left[ (Q\tilde{v}_x\Phi_\omega, \tilde{L}^{'-1}_{FF} Q \tilde{v}_y \Phi_{er} )_{ND} - (Q\tilde{v}_y\Phi_\omega, \tilde{L}^{'-1}_{FF} Q \tilde{v}_x \Phi_{er} )_{ND}  \right] \nonumber \\
\tilde{\Omega}^o_{\omega, T_r} &= - \frac{1}{2} \left[ (Q\tilde{v}_x\Phi_\omega, \tilde{L}^{'-1}_{FF} Q \tilde{v}_x \Phi_{er} )_{ND} + (Q\tilde{v}_y\Phi_\omega, \tilde{L}^{'-1}_{FF} Q \tilde{v}_y \Phi_{er} )_{ND}  \right] \nonumber \\
\tilde{\Omega}_{T_t,\omega} &= - \frac{1}{2} \left[ (Q\tilde{v}_x\Phi_{et}, \tilde{L}^{'-1}_{FF} Q \tilde{v}_y \Phi_\omega )_{ND} - (Q\tilde{v}_y\Phi_{et}, \tilde{L}^{'-1}_{FF} Q \tilde{v}_x \Phi_\omega )_{ND}  \right] \nonumber \\
\tilde{\Omega}^o_{T_t,\omega} &= - \frac{1}{2} \left[ (Q\tilde{v}_x\Phi_{et}, \tilde{L}^{'-1}_{FF} Q \tilde{v}_x \Phi_\omega )_{ND} + (Q\tilde{v}_y\Phi_{et}, \tilde{L}^{'-1}_{FF} Q \tilde{v}_y \Phi_\omega )_{ND}  \right] \nonumber \\
\tilde{\lambda}_{T_t,T_t} &= - \frac{1}{2} \left[ (Q\tilde{v}_x\Phi_{et}, \tilde{L}^{'-1}_{FF} Q \tilde{v}_x \Phi_{et} )_{ND} + (Q\tilde{v}_y\Phi_{et}, \tilde{L}^{'-1}_{FF} Q \tilde{v}_y \Phi_{et} )_{ND}  \right] \nonumber \\
\tilde{\lambda}^o_{T_t,T_t} &= - \frac{1}{2} \left[ (Q\tilde{v}_x\Phi_{et}, \tilde{L}^{'-1}_{FF} Q \tilde{v}_y \Phi_{et} )_{ND} - (Q\tilde{v}_y\Phi_{et}, \tilde{L}^{'-1}_{FF} Q \tilde{v}_x \Phi_{et} )_{ND}  \right] \nonumber \\
\tilde{\lambda}_{T_t,T_r} &= - \frac{1}{2} \left[ (Q\tilde{v}_x\Phi_{et}, \tilde{L}^{'-1}_{FF} Q \tilde{v}_x \Phi_{er} )_{ND} + (Q\tilde{v}_y\Phi_{et}, \tilde{L}^{'-1}_{FF} Q \tilde{v}_y \Phi_{er} )_{ND}  \right] \nonumber \\
\tilde{\lambda}^o_{T_t,T_r} &= - \frac{1}{2} \left[ (Q\tilde{v}_x\Phi_{et}, \tilde{L}^{'-1}_{FF} Q \tilde{v}_y \Phi_{er} )_{ND} - (Q\tilde{v}_y\Phi_{et}, \tilde{L}^{'-1}_{FF} Q \tilde{v}_x \Phi_{er} )_{ND}  \right] \nonumber \\
\tilde{\Omega}_{T_r,\omega} &= - \frac{1}{2} \left[ (Q\tilde{v}_x\Phi_{er}, \tilde{L}^{'-1}_{FF} Q \tilde{v}_y \Phi_\omega )_{ND} - (Q\tilde{v}_y\Phi_{er}, \tilde{L}^{'-1}_{FF} Q \tilde{v}_x \Phi_\omega )_{ND}  \right] \nonumber \\
\tilde{\Omega}^o_{T_r,\omega} &= - \frac{1}{2} \left[ (Q\tilde{v}_x\Phi_{er}, \tilde{L}^{'-1}_{FF} Q \tilde{v}_x \Phi_\omega )_{ND} + (Q\tilde{v}_y\Phi_{er}, \tilde{L}^{'-1}_{FF} Q \tilde{v}_y \Phi_\omega )_{ND}  \right] \nonumber \\
\tilde{\lambda}_{T_r,T_t} &= - \frac{1}{2} \left[ (Q\tilde{v}_x\Phi_{er}, \tilde{L}^{'-1}_{FF} Q \tilde{v}_x \Phi_{et} )_{ND} + (Q\tilde{v}_y\Phi_{er}, \tilde{L}^{'-1}_{FF} Q \tilde{v}_y \Phi_{et} )_{ND}  \right] \nonumber \\
\tilde{\lambda}^o_{T_r,T_t} &= - \frac{1}{2} \left[ (Q\tilde{v}_x\Phi_{er}, \tilde{L}^{'-1}_{FF} Q \tilde{v}_y \Phi_{et} )_{ND} - (Q\tilde{v}_y\Phi_{er}, \tilde{L}^{'-1}_{FF} Q \tilde{v}_x \Phi_{et} )_{ND}  \right] \nonumber \\
\tilde{\lambda}_{T_r,T_r} &= - \frac{1}{2} \left[ (Q\tilde{v}_x\Phi_{er}, \tilde{L}^{'-1}_{FF} Q \tilde{v}_x \Phi_{er} )_{ND} + (Q\tilde{v}_y\Phi_{er}, \tilde{L}^{'-1}_{FF} Q \tilde{v}_y \Phi_{er} )_{ND}  \right] \nonumber \\
\tilde{\lambda}^o_{T_r,T_r} &= - \frac{1}{2} \left[ (Q\tilde{v}_x\Phi_{er}, \tilde{L}^{'-1}_{FF} Q \tilde{v}_y \Phi_{er} )_{ND} - (Q\tilde{v}_y\Phi_{er}, \tilde{L}^{'-1}_{FF} Q \tilde{v}_x \Phi_{er} )_{ND}  \right].
\end{align}

\end{widetext}
\section{Computational Details}
\label{app:Computational_Details}
One of the main results of this paper was to numerically compute the transport and relaxation coefficients given in Eqs. \eqref{eq:transport_coef_inner_product} and \eqref{eq:relaxation_current_inner_product}. The main challenge of these computations is that the inner products contain the inverse of the Boltzmann operator projected to the fast space $L^{-1}_{FF}$, which is an integral operator acting on an infinite dimensional space and hence practically impossible to invert. To avoid this, we will write the Boltzmann operator $L$ in polynomial basis and truncate the space in finite order. This is a reasonable approximation as we expect that the higher order fast variables contribute less to the observable quantities. The numerical integrations for each matrix entry is computed via the Monte Carlo integration in Python Vegas library. Each integral is computed with 1500000 data points and 10 iterations.

\subsection{The Choice of Polynomial Basis}
\label{app:computational_procedure}
We choose Hermite polynomials for our basis, as the steady-state distribution we assume is Gaussian and they are orthonormal under the inner product,
\begin{equation}
    (H_i(x),H_j(y)) \equiv \int d x e^{-x^2} H_i(x)H_j(x) = \delta_{ij},
\end{equation}
which is the same as the non-dimensional inner product we define in Eq. \eqref{eq:non-dim_inner_prod}, given polynomials are real functions. For other steady-state distributions, other suitable bases can be chosen, or they can be numerically generated via the Gram-Schmidt procedure for numerical solutions.

The full form of the basis that we use in the velocity and angular velocity phase space is given as
\begin{equation}
    H_{ijk} \equiv H_{i}(\tilde{v}_x)H_j(\tilde{v}_y) H_k (\tilde{\omega} - \tilde{\omega}_0),
\end{equation}
where $H_i$ is the $i$th physicist's Hermite polynomial normalized with respect to the non-dimensional inner product $(H_i,H_j)_{ND} = \delta_{ij}$.
\subsection{Matrix Elements of the Boltzmann Operator $\boltzmannop$}
Using the basis and the inner product we chose, we can write down the matrix elements of the dimensionless Boltzmann operator $\tilde{\boltzmannop}$ \footnote{In fact, practically, computing the matrix elements of the Hermitian conjugate of the Boltzmann operator $\boltzmannop^\dagger$ is easier than computing the matrix elements of the Boltzmann operator $\boltzmannop$ itself. However, the description of the procedures below are equally applicable to both.} as
\begin{equation}
    \tilde{\boltzmannop}_{ijk,lmn} \equiv ( H_{ijk}, \tilde{\boltzmannop}[H_{lmn}]   )_{ND} = ( \tilde{\boltzmannop}^\dagger[H_{ijk}], H_{lmn}   )_{ND}.
\end{equation}
As we mentioned before, we will truncate the basis such that $n_{max} = 4$ and set $\tilde{\boltzmannop} \approx \sum_{ 0 \leq i,j,k,l,m,n \leq 4 } \tilde{\boltzmannop}_{ijk,lmn} |ijk)(lmn| $. With the map $i,j,k \rightarrow 25j + 5k + i $ between indices, $\tilde{\boltzmannop}$ can be stored as a 125x125 matrix. 

After we have the matrix form, we can generate the slow and fast spaces by observing the eigenvalue spectrum of the Hermitian conjugate of the Boltzmann operator $\tilde{\boltzmannop}^\dagger$. How these basis states are chosen is described in detail in Section \ref{hydrodynamic_variables}. Then $\tilde{\boltzmannop}_{FF}$ would be a smaller matrix defined as,
\begin{align}
    \tilde{\boltzmannop}_{FF} \approx \sum_{ i,j,k,l,m,n \not\in \mathcal{S}} \tilde{\boltzmannop}_{ijk,lmn}|ijk)(lmn|,
\end{align}
which is not a singular matrix anymore, and can be inverted to obtain an approximate expression for $\tilde{\boltzmannop}_{FF}^{-1}$. After that, all the terms that appear in Eqs. \eqref{eq:Boltzmann_hydro_mass} - \eqref{eq:adibatic_hydro_fin} can be numerically computed via dot products in this Hermite basis. 
\subsection{ Matrix Elements of $\tilde{L}_\omega$ }
\label{app:torque_term_herm_basis}
As it is discussed in the main text, the Boltzmann operator $L$ is split into two parts: the linearized collision operator $\linearcollop$ and the torque term $\torquecollop$. Although both of these terms can be numerically computed, we will show here that with the recursive relations of the Hermite polynomials, the matrix entries of the torque term $\torquecollop$ can be analytically calculated, given that the steady-state distribution is Gaussian.

In that case, the explicit form of the dimensionless torque term $\tilde{L}_\omega$ is,
\begin{equation}
    \tilde{L}_\omega =  \partial_{\tilde{\omega}} - 2(\tilde{\omega} - \tilde{\omega}_0). 
\end{equation}
Using the following identities,
\begin{subequations}
    \begin{equation}
    \frac{d}{dx} H_n(x) = \sqrt{2n} H_{n-1}(x),
\end{equation}
and
\begin{equation}
    2xH_n(x) = \sqrt{2(n+1)} H_{n+1}(x) + \sqrt{2n} H_{n-1}(x),
\end{equation}
\end{subequations}
the matrix elements of $\partial_{\tilde{\omega}}$ reads,
\begin{align}
    [\partial_{\tilde{\omega}}]_{ij} &= \int d^2 \tilde{v} d\tilde{\omega} e^{-\tilde{v}^2 - (\tilde{\omega} - \tilde{\omega}_0 )^2 } H_{i_1}(\tilde{v}_x) H_{i_2}(\tilde{v}_y)
    H_{i_3}( \tilde{\omega} - \tilde{\omega}_0  ) \nonumber \\
    &\times H_{j_1}(\tilde{v}_x) H_{j_2}(\tilde{v}_y)
    \partial_{\tilde{\omega}}H_{j_3}( \tilde{\omega} - \tilde{\omega}_0  ) \nonumber \\
    &= \delta_{i_1,j_1} \delta_{i_2,j_2} \delta_{i_3, j_3-1} \sqrt{2j_3}.
\end{align}
Similarly, the matrix elements of $2(\tilde{\omega} - \tilde{\omega}_0)$ reads,
\begin{align}
    2[\tilde{\omega}-\tilde{\omega}_0]_{ij} &= \int d^2 \tilde{v} d\tilde{\omega} e^{-\tilde{v}^2 - (\tilde{\omega} - \tilde{\omega}_0 )^2 } \nonumber \\ &\times H_{i_1}(\tilde{v}_x) H_{i_2}(\tilde{v}_y)
    H_{i_3}( \tilde{\omega} - \tilde{\omega}_0  ) \nonumber \\
    &\times H_{j_1}(\tilde{v}_x) H_{j_2}(\tilde{v}_y)
    2(\tilde{\omega}-\tilde{\omega}_0)H_{j_3}( \tilde{\omega} - \tilde{\omega}_0  ) \nonumber \\
    &= \delta_{i_1,j_1} \delta_{i_2,j_2} \nonumber \\
    &\times \left[ \delta_{i_3, j_3-1} \sqrt{2j_3} + \sqrt{2(j_3+1)} \delta_{i_3,j_3+1} \right],
\end{align}
and the full torque operator becomes
\begin{equation}
    \left[\tilde{L}_\omega \right]_{ij} = -\sqrt{2(j+1)} \delta_{i,j+1}.
\end{equation}

\subsection{Slow Variables in Hermite Basis}

To compute the inner products, one also needs the slow variables $\Phi_\alpha$ and $v^\mu \Phi_\alpha$ written in the chosen polynomial basis, which can always be done numerically. However, in our case we can obtain exact expressions of them in terms of the Hermite polynomials, which are,
\begin{widetext}

\begin{align*}
    \Phi_\rho &= N_0^3  H_0(\tilde{v}_x)H_0(\tilde{v}_y)H_0(\tilde{\omega} -\tilde{\omega}_0 )\\
    \Phi_{v,i} &= \sqrt{2}N_1N_0^2  H_1(\tilde{v}_i)H_0(\tilde{v}_j)H_0(\tilde{\omega} -\tilde{\omega}_0 ), i \neq j\\
    \Phi_{\omega} &= \sqrt{2}N_1N_0^2  H_1(\tilde{v}_x)H_0(\tilde{v}_y)H_1(\tilde{\omega} -\tilde{\omega}_0 )\\
     \Phi_{e,t} &= N_2N_0^2  H_2(\tilde{v}_x)H_0(\tilde{v}_y)H_0(\tilde{\omega} -\tilde{\omega}_0 ) + N[2]N[0]^2  H_0(\tilde{v}_x)H_2(\tilde{v}_y)H_0(\tilde{\omega} -\tilde{\omega}_0 )\\
     \Phi_{e,r} &= \sqrt{2} N_2N_0^2 | H_0(\tilde{v}_x)H_0(\tilde{v}_y) H_2( \tilde{\omega} - \tilde{\omega}_0 ) )\\
     \tilde{v}_x \Phi_m &= N_1N_0^2  H_1(\tilde{v}_x) H_0(\tilde{v}_y) H_0( \tilde{\omega} - \tilde{\omega}_0 )\\
     \tilde{v}_y \Phi_m &= N_1N_0^2  H_0(\tilde{v}_x) H_1(\tilde{v}_y) H_0( \tilde{\omega} - \tilde{\omega}_0 ) \\
     \tilde{v}_x \Phi_{v,x} &= \sqrt{2}\left[ N_2N_0^2 H_2(\tilde{v}_x) H_0(\tilde{v}_y) H_0( \tilde{\omega} - \tilde{\omega}_0 ) + N_0^3 H_0(\tilde{v}_x) H_0(\tilde{v}_y) H_0( \tilde{\omega} - \tilde{\omega}_0 )/2    \right]\\
     \tilde{v}_y \Phi_{v,x} &= \sqrt{2} N_1^2 N_0 H_1(\tilde{v}_x) H_1(\tilde{v}_y) H_0( \tilde{\omega} - \tilde{\omega}_0 )\\
     \tilde{v}_x \Phi_{et} &= N_3N_0^2 H_3(\tilde{v}_x) H_0(\tilde{v}_y) H_0( \tilde{\omega} - \tilde{\omega}_0 ) + N_1N_0^2 H_1(\tilde{v}_x) H_0(\tilde{v}_y) H_0( \tilde{\omega} - \tilde{\omega}_0 ) + N_2N_1N_0 H_1(\tilde{v}_x) H_2(\tilde{v}_y) H_0( \tilde{\omega} - \tilde{\omega}_0 )\\
     \tilde{v}_y \Phi_{et} &= N_3 N_0^2 H_0(\tilde{v}_x) H_3(\tilde{v}_y) H_0( \tilde{\omega} - \tilde{\omega}_0 ) + N_1 N_0^2 H_0(\tilde{v}_x) H_1(\tilde{v}_y) H_0( \tilde{\omega} - \tilde{\omega}_0 ) + N_2 N_1 N_0 H_2(\tilde{v}_x) H_1(\tilde{v}_y) H_0( \tilde{\omega} - \tilde{\omega}_0 )\\
     \tilde{v}_x \Phi_\omega &= \sqrt{2} N_1^2 N_0  H_1(\tilde{v}_x) H_0(\tilde{v}_y) H_1( \tilde{\omega} - \tilde{\omega}_0 )\\
     \tilde{v}_y \Phi_\omega &= \sqrt{2} N_1^2 N_0 H_0(\tilde{v}_x) H_1(\tilde{v}_y) H_1( \tilde{\omega} - \tilde{\omega}_0 )\\
     \tilde{v}_x \Phi_{e,r} &= \sqrt{2}N_2N_1N_0 H_1(\tilde{v}_x) H_0(\tilde{v}_y) H_2( \tilde{\omega} - \tilde{\omega}_0 )\\
     \tilde{v}_y \Phi_{e,r} &= \sqrt{2}N_2N_1N_0 H_0(\tilde{v}_x) H_1(\tilde{v}_y) H_2( \tilde{\omega} - \tilde{\omega}_0 ),
\end{align*}
\end{widetext}
where $N_n = \frac{\pi^{1/4} \sqrt{n!} }{\sqrt{2^n}}$ is the normalization coefficient for the monoidal Hermite polynomials (normalized in a way that the coefficient of the leading term is set to 1). With all of these combined, the calculation is reduced to simple inner products of finite dimensional vectors.

\subsection{Reducing the Dimensionality}

The most expensive part of the procedure described above is the calculation of the matrix elements of the linear collision operator $\tilde{\linearcollop}_{ij}$.  Each matrix entry is a 7-dimensional integral over the velocities and angular velocities of two colliding particles and the apse vector. Depending on the accuracy sought, this calculation can be very costly.

In this section we introduce an analytical trick to reduce the dimensionality from 7 to 4. In 3D system, same trick can be used to reduce it from 14 to 8. 

We first make the change of variables that we used to prove the rotational invariance: $\vec{u} \equiv \frac{\vec{c}_1 - \vec{c}_2}{\sqrt{2}}$, $ \vec{C} \equiv \frac{\vec{c}_1 + \vec{c}_2}{\sqrt{2}}$, $ \gamma \equiv \frac{\omega_1 - \omega_2}{\sqrt{2}} $ and $\Omega \equiv \frac{\omega_1 + \omega_2}{\sqrt{2}}$, where for notational convenience we dropped tildes, but all the variables are again dimensionless at this step. First, note that $v_1^2 + v_2^2 = u^2 + C^2$ and $ (\omega_1 - \omega_0)^2 + (\omega_2 - \omega_0)^2 = (\Omega - \sqrt{2}\omega_0)^2 + \gamma^2 $. However, the more important feature of these variables is that $\vec{C}$ and $\gamma$ are conserved during collisions. Moreover, $\vec{u}'$ and $\Omega'$ after collision do not depend on $\gamma$ or $\vec{C}$. To see how this helps, consider the matrix elements of the collision operator in these new variables, which are,
\begin{widetext}
\begin{align}
    \tilde{L}_{ij} = \frac{1}{\sqrt{2}} \int d^2 u d^2 C d\gamma d\Omega |\vec{u}\cdot\apsevector|\Theta(-\vec{u}\cdot\apsevector)e^{-u^2 - C^2 - \gamma^2 -(\Omega - \sqrt{2}\omega_0)^2 } H_{i_1}\left( \frac{u_x + C_x}{\sqrt{2}} \right) H_{i_2}\left( \frac{u_y + C_y}{\sqrt{2}} \right) H_{i_3}\left( \frac{ \Omega + \gamma}{\sqrt{2}} -\omega_0 \right) \nonumber \\
    \times \Bigg[ H_{j_1}\left( \frac{u'_x + C_x}{\sqrt{2}} \right) H_{j_2}\left( \frac{u'_y + C_y}{\sqrt{2}}  \right) H_{j_3}\left( \frac{\Omega' + \gamma}{\sqrt{2}} -\omega_0 \right) + H_{j_1}\left( \frac{-u'_x + C_x}{\sqrt{2}} \right) H_{j_2}\left( \frac{-u'_y + C_y}{\sqrt{2}}  \right) H_{j_3}\left( \frac{\Omega' - \gamma}{\sqrt{2}} -\omega_0 \right) \nonumber \\
    H_{j_1}\left( \frac{u_x + C_x}{\sqrt{2}} \right) H_{j_2}\left( \frac{u_y + C_y}{\sqrt{2}}  \right) H_{j_3}\left( \frac{\Omega + \gamma}{\sqrt{2}} -\omega_0 \right) + H_{j_1}\left( \frac{-u_x + C_x}{\sqrt{2}} \right) H_{j_2}\left( \frac{-u_y + C_y}{\sqrt{2}}  \right) H_{j_3}\left( \frac{\Omega - \gamma}{\sqrt{2}} -\omega_0 \right) \Bigg].
\end{align}
\end{widetext}
Now, since $u'_{x,y}$ and $\Omega'$ do not depend on $\vec{C}$ and $\gamma$, integrations over $C_x$, $C_y$ and $\gamma$ can be done beforehand. Say,
\begin{align}
    \int dC_x e^{-C_x^2} H_{i_1}\left( \frac{u_x + C_x }{\sqrt{2}} \right) H_{j_1}\left( \frac{u_x' + C_x}{\sqrt{2}} \right) \equiv P_{i_1,j_1}(u'_x, u_x),
\end{align}
can be done independently from the rest of the integration. Moreover, if we define this polynomial once, we can re-obtain all the separate integrals from it as,
\begin{widetext}
\begin{align}
   \int dC_{x,y} e^{-C_{x,y}^2} H_{i}\left( \frac{ \pm u_{x,y} + C_{x,y} }{\sqrt{2}} \right) H_{j}\left( \frac{\pm u_{x,y}' + C_{x,y}}{\sqrt{2}} \right)
   \equiv P_{i,j}(\pm u_{x,y},\pm u'_{x,y}, ),
\end{align}
and,
\begin{align}
    \int d\gamma H_i \left( \frac{\Omega-\sqrt{2}\omega_0 \pm \gamma}{\sqrt{2}} \right) H_j \left( \frac{\Omega'-\sqrt{2}\omega_0 \pm \gamma}{\sqrt{2}} \right) e^{-\gamma^2} \nonumber \\
    = (-1)^i (-1)^j \int d\gamma H_i \left( \frac{ \mp (\Omega-\sqrt{2}\omega_0) + \gamma}{\sqrt{2}} \right) H_j \left( \frac{ \mp (\Omega'-\sqrt{2}\omega_0) + \gamma}{\sqrt{2}} \right) e^{-\gamma^2} \equiv (-1)^i (-1)^j P_{i,j}( \mp \Omega, \mp \Omega'  ),
\end{align}
\end{widetext}
where we used the property of Hermite polynomials $H_n(x) = (-1)^nH_n(-x)$ on the second line. These integrals are simply Gaussian moments of $C_{x,y}$ and $\gamma$ and can be performed symbolically using Mathematica.

\subsection{Basis of irreducible representations}
\label{app:Irrep}
In this section, we will introduce a basis which makes the interpretation of the transport coefficients more intuitive. Consider the following basis of 2-tensors in 2 space dimensions:
\begin{subequations}
\begin{align}
      \tau^0 &= \begin{pmatrix}
        1 & 0\\
        0 & 1
    \end{pmatrix} 
    \qquad
    \tau^1 = \begin{pmatrix}
        0 & -1\\
        1 & 0
    \end{pmatrix}\\
    \tau^2 &= \begin{pmatrix}
        1 & 0\\
        0 & -1
    \end{pmatrix}
    \qquad
    \tau^3 = \begin{pmatrix}
        0 & 1\\
        1 & 0
    \end{pmatrix}
\end{align}
\end{subequations}
which are orthonormal under the inner product $\frac{1}{2}\tau^\alpha_{ij}\tau^\beta_{ij} = \delta_{\alpha\beta}$.
This basis is particularly convenient for the expression of the viscosity tensor as it decomposes the velocity gradients $\partial_i \vec{u}_j$ into volume expansion rate $ \dot{e}_0 \equiv \tau^0_{ij} \partial_i u_j$, rotation rate $\tau^1_{ij} \partial_i u_j$, and shear rates $\tau^{2,3}_{ij} \partial_i u_j$, respectively. Then the stress tensor in this basis represents the response to these deformation rates, and the viscosity tensor couples different deformation rates and stress responses. Although this representation does not give an intuitive interpretation for the rest of the transport coefficients, it also splits them into their parallel and transverse parts, as we demonstrate in the following.

We define the basis transformation for 2-tensors $\Lambda_{ij}$ as

\begin{equation}
    \Lambda^\alpha = \frac{1}{2} \tau^\alpha_{ij} \Lambda_{ij},
\end{equation}
and 4-tensors $\eta_{ijkl}$ as,
\begin{equation}
    \label{eq:visc_tensor_irrep}
    \eta^{\alpha \beta} = \frac{1}{4} \tau^{\alpha}_{ij} \eta_{ijkl} \tau^{\beta}_{kl}.
\end{equation}
Therefore, in the irreducible representation, the viscosity tensor $\eta^{\alpha \beta}$ becomes a 4 by 4 matrix while the rest of the transport coefficients become 4 dimensional vectors. The rotational invariance  impose further restrictions on these matrices and vectors, and for the viscosity we must have a matrix of the form,
\begin{equation}
    \eta_{\alpha \beta} = \begin{pmatrix}
        \eta_{00} & \eta_{01} & 0 & 0\\
        \eta_{10} & \eta_{11} & 0 & 0\\
        0 & 0 & \eta_{22} & \eta_{23}\\
        0 & 0 & -\eta_{23} & \eta_{22}
    \end{pmatrix},
\end{equation}
and the rest of the transport coefficients must be of the form,
\begin{equation}
    \Lambda^\alpha = \begin{pmatrix}
        \Lambda^0\\
        \Lambda^1\\
        0\\
        0
    \end{pmatrix}.
\end{equation}
The exact same procedure can be applied to relaxation coefficients, all of which will be 4D vectors in this representation. A more detailed discussion of the restrictions of rotational invariance can be found in the supplemental text of Ref. \cite{fruchart2023odd}.

\label{section:dimension_red}
\bibliography{ref}
\end{document}